\documentclass{emulateapj}
\usepackage[intlimits,fleqn]{amsmath}

\newcommand{\D}[1]{\, d #1 \,}
\newcommand{\vc}[1]{{\bf#1}}
\newcommand{\vch}[1]{\hat{\bf#1}}
\newcommand{\avg}[1]{\left< #1 \right>}
\newcommand{\ft}[1]{\widetilde{#1}}

\newcommand{\eps}{\varepsilon}

\newcommand{\sourcedir}{\texttt{www.astro.wisc.edu$/\!\!\sim$lazarian$/$simulations}}

\newcommand{\inctab}[1]{\includegraphics[scale=.26]{#1}}

\shorttitle{Turbulence Spectra from Spectral Lines}
\shortauthors{Chepurnov \& Lazarian}

\begin{document}


\title{Turbulence Spectra from Doppler-broadened Spectral Lines: Tests of the Velocity Channel Analysis and Velocity Coordinate Spectrum Techniques}

\author{A. Chepurnov, A. Lazarian}
\affil{Department of Astronomy, University of Wisconsin, Madison, USA}

\begin{abstract}
Turbulent motions induce Doppler shifts of observable emission and absorption lines motivating studies of turbulence using precision spectroscopy. We provide the numerical testing of the two most promising techniques, Velocity Channel Analysis (VCA) and Velocity Coordinate Spectrum (VCS). We obtain an expression for the shot noise that the discretization of the numerical data entails and successfully test it. We show that the numerical resolution required for recovering the underlying turbulent spectrum from observations depend on the spectral index of velocity fluctuations, which makes low resolution testing misleading. We demonstrate numerically that, dealing with absorption lines, sampling of turbulence along just a dozen directions provides a high quality spectrum with the VCS technique.
\end{abstract} 

\section{Introduction}

As a rule, astrophysical fluids are turbulent and the turbulence is magnetized. This ubiquitous turbulence determines the transport properties of the interstellar medium (see \citealt{EF96}, \citealt{Stu01}, \citealt{Bal06}) and the intracluster medium (see \citealt{Sun03}, \citealt{Ens06}, \citealt{Laz06b}), many properties of Solar and stellar winds (see \citealt{HMG80}), and so on. To understand heat conduction and the propagation of cosmic rays and electromagnetic radiation in different astrophysical environments it is absolutely essential to understand the properties of the underlying magnetized turbulence. The fascinating processes of star formation (see \citealt{MKT02}, \citealt{Elm02}, \citealt{MLK04}) and interstellar chemistry (see \cite{Fal06} and references therein) are also intimately related to properties of magnetized compressible turbulence (see reviews by \citealt{ES04}). 

We should stress that while the density fluctuations are an indirect way of testing turbulence, the most valuable information is given by the velocity field statistics encoded in spectrometric observations. The problem with those, that has been realized from the very start of the research in the field, is that the most\footnote{In some situations the data is more sparse than that. For instance, absorption lines usually do not provide good spatial sampling of the turbulent volume. Moreover, interferometric measurements of the emitting volumes may not have the entire u-v coverage to restore the spatial distribution of PPV emissivity either.}  that one can get from such observations are the Position-Position-Velocity (PPV) data cubes, where at the every point the image of the emitting turbulent volume the Doppler-broaderned spectrum is measured. To recover the statistics of turbulent velocity from the PPV, one has to account for the mapping of the turbulent velocity field from the real space to PPV data cubes. 

At the moment, several options for turbulence studies are available. For instance, using velocity centroids allows recovering of velocity spectrum (\citealt{EL05}, Esquivel et al. 2007). As shown there, this technique fails to recover the underlying velocity spectra for supersonic turbulence. For such turbulence one can use the Velocity Channel Analysis (VCA) (see \citealt{LP00}), which deals with channel maps or 2D slices of PPV, and Velocity Coordinate Spectrum (VCS) (see \citealt{LP00}, \citealt{LP06}, Lazarian \& Pogosyan 2008, henceforth, LP00, LP04 LP06, LP08, respectively, \citealt{LC08}, henceforth CL08) which studies the fluctuations of PPV intensities along the v-coordinate (see Lazarian 2008 for a review).

VCA has been already used with both atomic hydrogen and molecular line data to recover the underlying turbulence spectra (see review by \citealt{Laz06c}). The technique has been tested numerically for the optically thin case in \cite{Laz01} and \cite{Esq03} and also with absorption effects in \cite{Pad06}. However, rather strong shot-noise was observed in the synthetic channel maps. For instance, numerical simulations for VCA in \cite{Esq03} show that discretization over line of sight may substantially distort the spectrum at high wavenumbers. It was also shown for the thin-slice (velocity-dominated) case that the impact if shot-noise can be minimized by using thickest possible velocity slice. The shot-noise was also observed to decrease with the increase of the cube size used for analysis. Although the effect was associated with the discretization of the numerical data, a quantitative description of the noise was missing. Moreover, we feel that this effect confused some researchers who started to wonder about the practical utility of the technique \citep{Miv03}. 

While VCA is an established tool with numerous examples of practical applications, VCS is a more recent development. Although it was formulated in the same paper as VCA, i.e. LP00, the actual application of VCS is just beginning \citep{Che08}. 

In this paper we provide the numerical testing for both the VCA and the VCS techniques. In particular, we subject the origin of shot noise to scrutiny. 

In in Sect. \ref{sect:turb}, we describe general considerations about studied turbulent fields. In Sect. \ref{sect:bf}, we provide some facts about the VCA and VCS. In Sect. \ref{sect:sn}, we analyze shot-noise. In Sect. \ref{sect:res}, we describe the results of numerical testing. The case of observations in absorption lines is discussed in Sect. \ref{sect:sal}. We provide  a discussion in Sect. \ref{sect:disc}.

\section{Underlying fluctuations of velocity and density} \label{sect:turb}

Our assumptions about underlying turbulence coincide with those in LP00. In particular, in what follows we assume that turbulent velocity and density are homogeneous\footnote{Random fields are homogeneous in the statistical sense when their statistical properties do not depend on the position. This allows us to keep only the dependencies of our correlation functions on the distance between the points. Homogeneity is also a requirement for providing the volume averaging that observers would rely on to apply our results} and isotropic and admit a statistical description in terms of correlation (or structure) functions and the corresponding power spectra\footnote{The latter pair of functions is related to each other through Fourier transform (see Monin \& Yaglom 1975).}. The assumption of velocity isotropy and its relevance to the actual magnetized astrophysical fluids was discussed at length in LP00, Lazarian \& Pogosyan (2004), LP06. There it was argued that the degree of anisotropy that is induced by the interstellar magnetic field does not compromise the power spectrum recovered from observations. This argument was supported by the analysis of synthetic maps obtained using 3D MHD simulations of turbulence (Esquivel et al. 2003).

For a vector field like velocity, the correlation function and the correspondent power spectrum take the form of a tensor. In addition, the power spectrum is different for solenoidal and potential field components. For each case the tensor properties can be represented by a factor that depends only on angles. Additional scalar factor is responsible for energy distribution in wavenumber space.

Because of the self-similarity of turbulence, we expect a power law scaling for the power spectrum in the inertial range. This power law can be either ``long wave dominated'' (or ``steep''), if the respective spectral index $\alpha_{\epsilon}$ is greater than $3$, or ``short wave dominated'' (or ``shallow'') otherwise (see LP00). To keep total energy limited, both types of spectrum must have a cutoff: at low wavenumbers for the steep one, and at high wavenumbers for the shallow one. The cutoff is a physical requirement that is satisfied in all astrophysical systems.

In this paper we assume the velocity spectrum to be steep, as a shallow velocity spectrum is not physically motivated (see Cho \& Lazarian 2005). For the density spectrum we consider both steep and shallow spectra. The shallow spectrum for density emerges for high Mach numbers (see Padoan et al. 2004, Bereznyak, Lazarian \& Cho 2005, Kim \& Rye 2005, Kowal, Lazarian \& Beresnyak 2007). 

The quantities we deal with in spectral line observations are the velocity and the gas emissivity. The latter can be proportional to the density of gas $\rho$ (e.g. emission lines of HI) or $\rho^2$ (e.g. recombination lines in plasma). The latter regime modifies the analysis. Namely, for steep density spectrum (e.g. when $\alpha_{\rho}>3$), the asymptotic slope of the emissivity spectrum is the same, and different, $\alpha_\eps=2\alpha_\rho-3$, for the shallow density spectrum. 

We are interested in velocity statistics, and the gas emissivity is considered only because its fluctuations also affect the observable PPV fluctuations. The influence of the fluctuations of gas emissivity is mitigated when they have a steep spectrum. In this situation, in fact, they could be ignored within the VCS technique. However, when fluctuations of gas emissivity follow shallow spectrum these fluctuations should be properly accounted for.

\section{Basics of VCA \& VCS} \label{sect:bf}

\begin{figure}
\begin{center}
\plotone{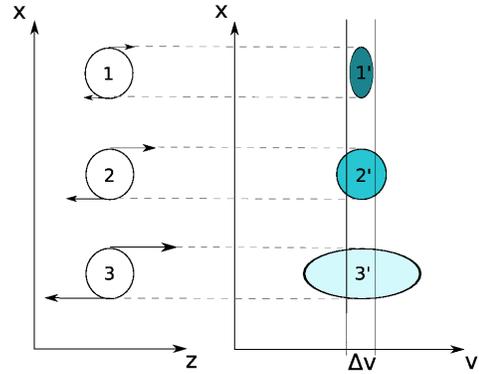}
\end{center}
\caption{An illustration of the mapping from the real space to the PPV space. In the real space 3 eddies above have the same spatial size, but different velocities. They are being mapped to the PPV space and there they have the same PP dimensions, but a different V-size. The larger is the velocity of eddies, the larger the V-extend of the eddies, the less density of atoms over the image of the eddy. For the eddy 3 the slice is "thin", i.e. the statistics is defined by the velocity spectrum. The images of eddies corresponding to velocities less than the channel thickness is different. The velocity slice is "thick" for the eddy 1, i.e. the statistics is defined by the emissivity spectrum, and the eddy 2 is in an intermediate regime.\label{fig:VCA_res}}
\end{figure}

Fig. \ref{fig:VCA_res} illustrates the essence of the VCA technique. The data in the PPV data cube are analyzed using slices of $\Delta V$ thickness. This corresponds to the analysis of intensity fluctuations in channel maps (see Green 1993), which predates the technique. The gist of the technique is a relation between the variations of the spectral index of channel map fluctuations and the thickness of the channel maps. The differences in the statistics of the fluctuations stem from the fact that the images of the eddies can have an extent both larger and smaller than PPV slice thickness $\Delta V$. The former case is termed by LP00  ``thin slice'', and the latter is termed ``thick slice''.


\begin{figure}
\begin{center}
\plotone{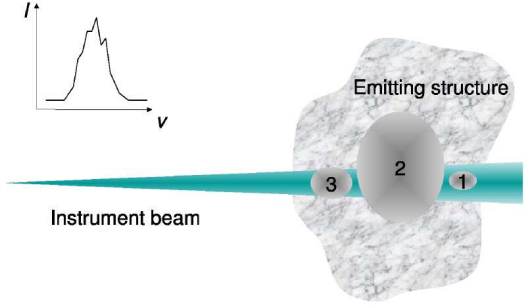}
\end{center}
\caption{VCS technique: effects of resolution. The fluctuations {\it along} the velocity coordinate are analyzed. Eddies within the telescope size beam, e.g. eddy  1, are in a low resolution mode. Eddies with the size exceeding the one of the beam, e.g. eddy 2, are in the high resolution mode.\label{fig:VCS_res}} 
\end{figure}

In Fig. \ref{fig:VCS_res} we illustrate the essence of the VCS technique. Depending on the resolution of the telescope one may or may not resolve the spatial extent of the eddies under study. This  also results in two distinct regimes of turbulence studies. The ability to study turbulence in the turbulent volumes that are not spatially resolved is a unique ability of VCS. 


Below we briefly overview the main analytical results, obtained in LP00, LP04, LP06 and CL08, which are relevant for this paper. We, however, do not deal with the case of VCA/VCS studies of turbulent volumes where self-absorption is important\footnote{In this case analysis is more simple if we use Fourier space. Absorption effects can be described only in real space, which implies the use of structure functions instead (see \cite{LP04}, \cite{LP06}). It is shown in \cite{LP04} that for VCA absorption can be neglected as soon as the scale of turbulence under study is less than the physical length corresponding to the optical depth of unity. A similar criterion is applicable to VCS \citep{LP06}.} (LP04) or the case of saturated absorption lines (LP08). In terms of the presentation of our results, below, we use the approach and notations from CL08. 

Our calculations are based on the following expression for a spectral line signal, measured at velocity $v_0$ and given beam position $\vch{e}$ (see Appendix~\ref{sect:sig} for the details): 
\begin{equation} \label{eq:sig_rat}
S(\vch{e},v_0)
  = \int w(\vch{e},\vc{r}) \D{\vc{r}} \eps(\vc{r}) f(v_r(\vc{r}) + v_r^{reg}(\vc{r}) - v_0).
\end{equation}
where $\eps$ is normalized emissivity, $v_r^{reg}$ is a line-of-sight component of the regular velocity (e.g. the velocity arising from the galactic velocity shift), $v_r$ is the line-of-sight component of the random turbulent velocity, $f$ denotes the convolution between channel sensitivity function\footnote{I.e. amplitude-frequency response of the channel normalized to the integral value with frequency in velocity units} and Maxwellian distribution of velocities of gas particles, defined by temperature of emitting medium, and $w$ is a window function defined as follows: 
\begin{equation} \label{eq:w}
w(\vch{e},\vc{r})
  \equiv \frac{1}{r^2} w_b(\vch{e},\vch{r}) w_\eps(\vc{r}),
\end{equation}
where $w_b$ is an instrument beam, which depends on angular coordinates $\phi$ and $\theta$, while $w_\eps$ is a window function defining the extent of the observed object. For instance, for a turbulent volume confined by two planes perpendicular to the line of sight, the window function is as follows:
\begin{equation} 
w_\eps(z) =
  \left\{ 
    \begin{array}{ll} 
      1, & z \in [z_0,z_1] \\
      0, & z \notin [z_0,z_1] \\
    \end{array} 
  \right.
\end{equation} 

In what follows, we consider the Fourier transform of a spectral line\footnote{Variable $k_v$ plays here the role of $k_z$ in LP00, being, however, different in dimension ($k_z = b k_v$, see Eq. (\ref{eq:uz_lin}). We use it here to avoid complications when $b=0$.}:

\begin{equation} \label{eq:sig_ft}
\begin{array}{ll}
\ft{S}(\vch{e},k_v) 
  & \equiv \frac{1}{2\pi} \int_{-\infty}^\infty S(v_0) e^{-i k_v v_0} \D{v_0} \\
  & = \ft{f}(k_v) \int w(\vch{e},\vc{r}) \D{\vc{r}} \cdot \\
  &   \eps(\vc{r}) \exp(-i k_v (v_r(\vc{r}) + v_r^{reg}(\vc{r}))).
\end{array}
\end{equation}
This function can be easily determined from observational data.

If we correlate $\ft{S}$ taken in two directions $1$ and $2$, we get the following measure, which can be used as a starting point for the mathematical formulation of both the VCA and VCS techniques (see CL08 for details):
\begin{equation} \label{eq:K12_def} 
\begin{array}{ll}
K(\vch{e}_1,\vch{e}_2,k_v) 
  & \equiv \avg{\ft{S}(\vch{e}_1,k_v) \ft{S}^*(\vch{e}_2,k_v)} \\ 
  & =\ft{f}^2(k_v) \int w(\vch{e}_1,\vc{r}) \D{\vc{r}} \int w(\vch{e}_2,\vc{r'}) \D{\vc{r'}} \cdot\\
  & \avg{\eps(\vc{r})\eps(\vc{r'})} \avg{\exp(-i k_v (v_r(\vc{r})-v_{r'}(\vc{r'})))}  \cdot\\
  & \exp(-i k_v (v_r^{reg}(\vc{r})-v_{r'}^{reg}(\vc{r'}))),
\end{array}
\end{equation} 
The first averaging gives us an emissivity correlation function $C_\eps(\vc{r}-\vc{r'})$. Averaging of the exponent can be performed with the assumption that the velocity statistics is Gaussian\footnote{We assume that the velocity field has a Gaussian Probability Distribution Function (PDF). The latter corresponds well to both experimental (Monin \& Yaglom 1976) and numerical (Biskamp 2003) data.}:
\begin{equation} 
\begin{array}{ll}
&\avg{\exp(-i k_v (v_r(\vc{r})-v_{r'}(\vc{r'})))} \\
&  = \exp\left(-\frac{k_v^2}{2} \avg{(v_r(\vc{r})-v_{r'}(\vc{r'}))^2}\right).
\end{array}
\end{equation}
To proceed, we assume that the beam separation and the beam width are both small enough that we can neglect the difference between $v_r$ and $v_z$ (we consider $z$-axis to be a bisector of the angle between beams). We also assume that $v_z^{reg}(\vc{r})$ depends only on $z$ and admits a linear approximation:
\begin{equation} \label{eq:uz_lin} 
v_z^{reg}(z)
  = b(z-z_0)+v_{z,0}^{reg},
\end{equation}
where $b$ characterizes regular velocity shear.

This leads us to
\begin{equation} \label{eq:K12_inter} 
\begin{array}{ll}
K(\vch{e}_1,\vch{e}_2,k_v)
  & =\ft{f}^2(k_v) \cdot \\
  & \int w(\vch{e}_1,\vc{r}) \D{\vc{r}} \int w(\vch{e}_2,\vc{r'}) \D{\vc{r'}} C_\eps(\vc{r}-\vc{r'}) \cdot \\
  & \exp \left(-\frac{k_v^2}{2} D_{vz}(\vc{r}-\vc{r'}) - ik_v b(z-z') \right), 
\end{array}
\end{equation}

Having substituted $\vc{r}-\vc{r'}$, we can write the following expression for $K$:
\begin{equation} \label{eq:K12} 
\begin{array}{ll}
K(\vch{e}_1,\vch{e}_2,k_v) 
  & =\ft{f}^2(k_v) \int g(\vch{e}_1,\vch{e}_2,\vc{r}) \D{\vc{r}} \cdot\\
  & C_\eps(\vc{r}) \exp \left(-\frac{k_v^2}{2} D_{vz}(\vc{r}) - ik_v b z \right), 
\end{array}
\end{equation}
where 
\begin{equation}
g(\vch{e}_1,\vch{e}_2,\vc{r}) 
  \equiv \int w(\vch{e}_1,\vc{r'})w(\vch{e}_2,\vc{r'}+\vc{r}) \D{\vc{r'}}, 
\end{equation} 
$\ft{f}$ is a Fourier transform of an effective channel sensitivity function $f$, $C_{\eps}$ is an emissivity correlation function, and $D_{vz}$ is a velocity structure tensor projection:
\begin{equation}
D_{vz}(\vc{r}-\vc{r'})
  \equiv \avg{(v_z(\vc{r})-v_z(\vc{r'}))^2}, 
\end{equation}

If we set $g=w_{b,a}(\vc{R}-\vc{R}_{12})w_{\eps,a}(z)$, where $\vc{R}_{12}$ is a beam separation in pictorial plane\footnote{The pictorial plane is the plane perpendicular to the line of sight. It provides us the coordinates in spatial units for the beam, if we consider beam's cross-section by a pictorial plane at the distance of a remote object.}, $w_{b,a}$ and $w_{\eps,a}$ are auto-convolutions of $w_b$ and $w_\eps$, which corresponds to the case of remote emitting object, we have for $K$:
\begin{equation} \label{eq:K12_R12} 
\begin{array}{ll}
K(\vc{R}_{12},k_v) 
  & =\ft{f}^2(k_v) \int w_{b,a}(\vc{R}-\vc{R}_{12})w_{\eps,a}(z) \D{\vc{r}} \cdot \\
  & C_\eps(\vc{r}) \exp \left(-\frac{k_v^2}{2} D_{vz}(\vc{r}) - i k_v b z \right) \\
  & =\ft{f}^2(k_v) \int \ft{w}_b^2(\vc{K}) \D{\vc{K}} \cdot \\
  & \int w_{\eps,a}(z) \D{\vc{r}} e^{-i\vc{K}(\vc{R}-\vc{R}_{12})} \cdot \\
  & C_\eps(\vc{r}) \exp \left(-\frac{k_v^2}{2} D_{vz}(\vc{r}) - i k_v b z \right)
\end{array}
\end{equation}
Taking a Fourier transform over $\vc{R}_{12}$, we have the power spectrum, designated in VCA as $P_s$: 
\begin{equation} \label{eq:Ps}
\begin{array}{ll}
P_s(\vc{K},k_v)
  & =\ft{f}^2(k_v) \ft{w_b}^2(\vc{K}) \int w_{\eps,a} \D{\vc{r}} \cdot \\
  & e^{-i(\vc{K}\vc{R}+b k_v z)}
    C_\eps(\vc{r}) \exp \left(-\frac{k_v^2}{2} D_{vz}(\vc{r}) \right),
\end{array}
\end{equation}
which coincides with the corresponding expression in LP00.

It is convenient to introduce $P_3$ as a 3D power spectrum of a \textit{PPV data cube} with assumption of infinite resolution over $v$ and $\vc{R}$:
\begin{equation} \label{eq:P3}
\begin{array}{ll}
P_3(\vc{K},k_v)
  & \equiv \avg{|\ft{S}(\vc{K},k_v)|^2}
  = \int w_{\eps,a} \D{\vc{r}} \cdot \\
  & e^{-i(\vc{K}\vc{R}+b k_v z)} C_\eps(\vc{r}) \exp \left(-\frac{k_v^2}{2} D_{vz}(\vc{r}) \right).
\end{array}
\end{equation}
In this case it is easy to see that $P_s(\vc{K},k_v) = \ft{f}^2(k_v) \ft{w_b}^2(\vc{K}) P_3(\vc{K},k_v)$.

VCA studies the two-dimensional spectrum of fluctuations within a velocity slice of a PPV cube:
\begin{equation} \label{eq:P2}
\begin{array}{ll}
P_2(\vc{K})
  &\equiv \int_{-\infty}^\infty \D{k_v} P_s(\vc{K},k_v) \\
  &= \ft{w_b}^2(\vc{K}) \int_{-\infty}^\infty \D{k_v} \ft{f}^2(k_v) P_3(\vc{K},k_v),
\end{array}
\end{equation}
which can be easily determined from observations. This technique provides a way to determine the slopes of velocity and density power spectra, $\alpha_v$  and $\alpha_\eps$. In fact, for a channel with small enough effective width\footnote{It is approximately equal to $\sqrt(\delta v_{ch}^2+(2 v_T)^2)$, where $\delta v_{ch}$ is the channel width and $v_T$ is the thermal velocity of emitting atoms.} (see Fig. \ref{fig:VCA_res}), $P_2$ is in the \textit{velocity dominated} regime with slope $(9-\alpha_v)/2$, provided that the density has a steep spectrum, i.e. $\alpha_\eps>3$. 

Whether or not the latter is true can be established through using from column density maps. 
As we neglect the effects of self-absorption, the column densities can be obtained via v-integration of PPV cubes.  Naturally, in the column density maps the spectrum is affected only by density and its slope is $\alpha_\eps$. The situation is a bit more complicated when the density is shallow (i.e. when $\alpha_\eps<3$), which is a case of high Mach number turbulence (see Beresnyak, Lazarian \& Cho 2006) and the density combines with velocity to affect the fluctuations in thin channels. For a more detailed analysis, see LP00. The predictions for different cases in the velocity-dominated regime are summarized in Table \ref{tab:P2_slopes}. Note that these results are obtained by assuming that the turbulent object under study is so distant that the geometry of parallel lines of sight is applicable.


\begin{deluxetable}{lll}
\tablecaption{VCA predictions about $P_2$ spectral index, steep density \label{tab:P2_slopes}}
\tablehead{
  \colhead{density spectrum} & 
  \colhead{2-d spectrum} & 
  \colhead{1-d spectrum}
}
\startdata
steep
&
\begin{math} 
  \frac{9-\alpha_v}{2}
\end{math}
&
\begin{math} 
  \frac{7-\alpha_v}{2} 
\end{math}
\\
shallow
&
\begin{math} 
  \frac{2\alpha_\eps-\alpha_v+3}{2}
\end{math}
&
\begin{math} 
  \frac{2\alpha_\eps-\alpha_v+1}{2}
\end{math}
\\
\enddata
\tablecomments{Here $\alpha_v$ is the velocity spectral index, $\alpha_\eps$ is the density spectral index. Density spectrum is considered to be ``steep'' if $\alpha_\eps > 3$ and ``shallow'' otherwise.}
\end{deluxetable}


The one-dimensional spectrum $P_1$ is the subject of the VCS studies. It corresponds to the case when the two beams involved in $K$ coincide, i.e. 
\begin{equation} \label{eq:P1_orig} 
\begin{array}{ll}
P_1(k_v)
  & \equiv K(\vc{R}_{12},k_v)\vert_{\vc{R}_{12}=0} 
  = \ft{f}^2(k_v) \int g(\vc{r}) \D{\vc{r}} \cdot \\
  & C_\eps(\vc{r}) \exp \left(-\frac{k_v^2}{2} D_{vz}(\vc{r}) - i k_v b z \right),
\end{array}
\end{equation}
which can be written also in terms of $P_3$ as:
\begin{equation} \label{eq:P1}
P_1(k_v)
  = \ft{f}^2(k_v) \int \D{\vc{K}} \ft{w_b}^2(\vc{K}) P_3(\vc{K},k_v).
\end{equation}
As it was shown in \cite{LP06} and \cite{LC08}, here we also have two spectral regimes. They depend on beamwidth (see Fig. \ref{fig:VCS_res} for illustration). For the \textit{high resolution mode} ($k_v$ is less than the velocity variance on the beam scale) the slope of $P_1$ is $2/(\alpha_v - 3)$, otherwise, in the \textit{low resolution mode} it is $6/(\alpha_v - 3)$ (a steep density spectrum is assumed for the both cases). A more complete list of the predictions for $P_1$ is presented in Table \ref{tab:P1_slopes}. The flat beam there corresponds to the telescope beam for which the resolution along one of the spatial axes is infinitely better than the resolution for the perpendicular axis. 


\begin{deluxetable}{llll}
\tablecaption{VCS predictions about $P_1$ spectral index, parallel lines of sight \label{tab:P1_slopes}}
\tablehead{
  \colhead{density spectrum} & 
  \colhead{pencil beam} & 
  \colhead{flat beam} & 
  \colhead{low resolution}
}
\startdata
steep
&
\begin{math} 
  \frac{2}{\alpha_v-3}
\end{math}
&
\begin{math} 
  \frac{4}{\alpha_v-3} 
\end{math}
&
\begin{math}
  \frac{6}{\alpha_v-3}
\end{math}
\\
shallow
&
\begin{math} 
  \frac{2(\alpha_\eps-2)}{\alpha_v-3} 
\end{math}
&
\begin{math} 
  \frac{2(\alpha_\eps-1)}{\alpha_v-3}
\end{math}
&
\begin{math}
  \frac{2 \alpha_\eps}{\alpha_v-3}
\end{math}
\\
\enddata
\tablecomments{Here $\alpha_v$ is the velocity spectral index, $\alpha_\eps$ is the density spectral index. Density spectrum is considered to be ``steep'' if $\alpha_\eps > 3$ and ``shallow'' otherwise.}
\end{deluxetable}


If we compare Eqs. (\ref{eq:P2}) and (\ref{eq:P1}), we see that the VCA and the VCS provide complimentary approaches and are related to the same quantity, $P_3$. Their properties must therefore be closely related.

\section{Shot-noise arising from numerical finite resolution} \label{sect:sn}

Earlier attempts to test the accuracy of VCA faced a problem; namely the distortion of the spectrum of fluctuations within a v-slice of the PPV data cube (see Lazarian et al. 2001). The effect from the very beginning was associated with the discretization of the numerical data, but attempts to remove the noise by the interpolation of the numerical data were not successful. In particular, in \cite{Esq03} it was shown that the quality of the spectrum in the v-slice depends on the number of points along the line of sight in the numerical cube used to create the PPV cube. The interference noise was termed \textit{shot-noise} to reflect its stochastic nature: insufficient statistics for the signal in spectrometer channels\footnote{If we take too low $N_z$ in our simulations, there can be too few points that contribute to the signal in a particular channel, which makes such a value ``unreliable''. However, one can face similar situation in X-ray observations, where individual photons are counted. In this case the time of  observation plays the role of $N_z$}. Below we show that the same type of noise is present in the VCS analysis. In this case, $P_1$ asymptotically approaches a constant minimum value for large $k_v$, which diminishes the spatial extent over which one can study turbulence (see Fig. \ref{fig:VCA_VCS_3_67_unfilt}, right).

To have a realistic spectral line we need a velocity field to be adequately resolved over line of sight (see also \citealt{IS03}). This means that the typical scale over $z$ for which velocity stays within a single channel near its extremum, should have enough discretization points, because such extrema give the dominant contribution to a spectrometer signal. 

Let us discuss the required resolution, related to the size $N_z$ of the data cube along z-axis. One can consider the non-linear space-velocity mapping\footnote{Such mapping can be illustrated by an example provided on Fig. \ref{fig:VCA_res}} induced by the turbulent velocity. The relation between the scale of real space $l$ and the corresponding velocity scale $v_l$ is given by $v_l\sim l^{(\alpha_v-3)/2}$, which for the Kolmogorov turbulence, i.e. $\alpha=11/3$, gives the familiar $v_l\sim l^{1/3}$ relation. The minimal scale $l_{min}$ which can be studied with either VCS or VCA technique corresponds to the distance between adjacent points in the original data cube. This minimal distance translates into the smallest velocity over which the velocity fluctuations can be studied, i.e. $v_{z,min}\sim l_{min}^{(\alpha_v-3)/2}$. This minimal velocity corresponds to the maximal wavenumber along the velocity coordinate, which can be recovered from the simulations $k_{v, max}\sim 1/v_{z, min}$. Therefore $k_{v, max}\sim  l_{min}^{-(\alpha-3)/2}$. If the data cube has $N_z$ points, $l_{min}\sim N_{z}^{-1}$. This results in $k_{v, max}\sim N_z^{(\alpha-3)/2}$. For $k_v$ numbers to contain the information about turbulence $k_v$ should be less than $k_{v, max}$. Consequently the number of points along the line of sight in the numerical data cube should scale as $ k_{v, max}^{2/(\alpha_v-3)}$. This provides the same dependence as the more detailed Eq.~(\ref{eq:Nz}).  

Naturally no VCA studies are meaningful below the channel thickness of $1/k_{v, max}$.

The arguments above also mean that, according to Table~2, the required number of points for flat beam scales as $k_{v, max}^{4/(\alpha_v-3)}$ and for low resolution as $k_{v, max}^{6/(\alpha_v-3)}$, which provides even steeper dependencies than the pencil beam. Thus, testing our formulae with low resolution will be more challenging.

To estimate the exact resolution dependence, we have to relate the correspondent typical scale over line of sight with the channel width. The natural choice is to calculate the velocity dispersion for wavelengths shorter than this scale. As mentioned above, the contribution of the velocity profile shapes other than extrema, provided by the remaining lower harmonics, can be neglected. Therefore, the scale found by reverting this expression is what we seek.

In other words, we divide the 1D velocity spectrum domain into the intervals separated by the scale in question $L_\Delta$, and, as spoken, assume that the harmonics in its low-wavenumber part provide an extremum of velocity. For the higher harmonics, in order to place the bulk of the signal inside one channel, we should demand that the correspondent variance $\sigma(L_\Delta)$ is about half a channel width:
\begin{equation} \label{eq:sigmabalance}
\sigma(L_\Delta) 
  \approx \frac{\sigma(\infty)}{2 N_\sigma},
\end{equation} 
where $L_\Delta$ is a scale at which velocity stays within a single channel near its extremum, $N_\sigma$ is the number of channels per velocity variance, and $\sigma(L)$ is a velocity variance for scales smaller than $L$, expressed through 1D power spectrum\footnote{This Should not be confused with $P_1$, which refers to PPV space} $E(k)$: 
\begin{equation} \label{eq:sigma}
\sigma^2(L)
  = 2 \int_{\frac{2\pi}{L}}^\infty E(k) \D{k}.
\end{equation} 

If a 3D power spectrum has a cutoff at $k_0=2\pi/L$, where $L$ is an injection scale, $E(k)$ has the following form for the isotropic 3D spectrum of $v_z$ 
\begin{equation} \label{eq:P_1d}
E(k)
  \sim \frac{1}{\alpha_v-2} \left\{ 
    \begin{array}{ll} 
      k^{-(\alpha_v-2)},   & |k| > k_0 \\
      k_0^{-(\alpha_v-2)}, & |k| \le k_0 \\
    \end{array} 
  \right.
\end{equation} 
Then, having solved Eq. (\ref{eq:sigmabalance}) for $L_\Delta$ we get
\begin{equation} \label{eq:L_delta}
L_\Delta 
  \approx \left(\frac{2 N_\sigma}{\sqrt{\alpha_v-2}}\right)^{-\frac{2}{\alpha_v-3}} L~.
\end{equation} 

If $n_\Delta$ is a number of points per $L_\Delta$, then the total number of points along $z$ is as follows: 
\begin{equation} \label{eq:Nz}
N_z
  \approx n_\Delta \frac{L_s}{L} \left(\frac{2 N_\sigma}{\sqrt{\alpha_v-2}}\right)^{\frac{2}{\alpha_v-3}} 
\end{equation} 
where $L_s$ is the size of emitting structure, $n_\Delta$ is determined from simulations and is experimentally found to be around 6.

The requirements for $N_z$ are rather demanding. If we take our typical settings: 128 channels per triple line width and $L_s/L=2$, then for $\alpha_v = 4$, $N_z$ should be at least $1.1\cdot10^4$ points, for $\alpha_v = 11/3$ the corresponding number is $4.3\cdot10^5$, while for $\alpha_v = 3.5$ the number is as big as $1.7\cdot10^7$. Therefore it is not surprising that without this knowledge some researchers could get puzzled by their results (e.g. \citealt{Miv03}).

Condition (\ref{eq:Nz}) guarantees that the spectrum is clear from shot-noise for all available $k_v$'s, up to $2\pi/\Delta v$, where $\Delta v$ is the channel width. This estimation refers to the case without smoothing along angular coordinates, i.e. for the high-resolution regime. The low-resolution VCS spectrum drops faster and therefore the requirements are even more restrictive.

\section{Results of Numerical Testing} \label{sect:res}

\subsection{Recovering Spectral Indexes}

VCS spectra, illustrating the condition given by Eq.~(\ref{eq:Nz}) are presented in Fig. \ref{fig:Nz_req}. The velocity spectral index $\alpha_v$ decreases from top to bottom (the values are 4, 3.78 and 3.56). According to our estimates above, this requires a drastic increase of the number of points. Indeed, $N_z$ is, respectively, $2^{14}$, $2^{17}$ and $2^{23}$. In the left column, with shot-noise clearly present, $N_z$ is 4 times undervalued.
Examples of restoration of the velocity spectral indices from simulations are presented in Tab. \ref{tab:alpha_v_re}.
They show good agreement. Values of $N_z$ have been chosen regarding Eq. (\ref{eq:Nz}).
Thus, knowing the requirements on $N_z$, we can generate the VCS and VCA spectra (see Fig. \ref{fig:VCA_VCS_4_00}, with $\alpha_v=4$). 


\begin{deluxetable}{lll}
\tablecaption{Restoring of velocity spectral index $\alpha_v$ from simulations \label{tab:alpha_v_re}}
\tablehead{
  \colhead{expected $\alpha_v$} & 
  \colhead{restored $\alpha_v$} & 
  \colhead{Nz} 
}
\startdata
3.56 & 3.54$\pm$0.01 & 8388608 \\
3.67 & 3.65$\pm$0.02 & 524288 \\
3.78 & 3.79$\pm$0.02 & 131072 \\
3.89 & 3.89$\pm$0.02 & 32768 \\
4.00 & 3.91$\pm$0.04 & 16384 \\
\enddata
\tablecomments{16 realizations of spectral line have been used for each $P_1$}
\end{deluxetable}


But for shallower velocity spectra, satisfying such conditions for VCA may be difficult, even if we use 2-d data instead of a 3-d data cube. For example, the condition (\ref{eq:Nz}) for the VCA simulations at $\alpha_v=11/3$ implies a square array with side $420000$, which is difficult to implement at present. If we decrease this size, the VCA spectrum gets distorted as seen in Fig. \ref{fig:VCA_VCS_3_67_unfilt}, left. Therefore instead of prescribing the number of VCA channels we have to seek alternative approaches.

\subsection{Choosing Optimal Thickness of Slices}

In \cite{Esq03} it was shown, that if we "optimize" the thickness of the velocity slice (which is equivalent to changing the velocity resolution), we get a correct spectrum. Let us consider why this can happen. Looking at the right panel on Fig. \ref{fig:VCA_VCS_3_67_unfilt}, we see that at higher $k_v$ the VCS spectrum is affected by shot-noise. As both VCA and VCS spectra are just different projections of the same 3-d power spectrum in the PPV space (see Eqs. (\ref{eq:P2}) and (\ref{eq:P1})), the noisy harmonics seen in the $P_1$ plot are most likely to affect the $P_2$ spectrum. If we use only $k_v$'s, which correspond to the intact part of $P_1$, the picture improves significantly, as seen in Fig. \ref{fig:VCA_VCS_3_67_filt}, left. Such truncation of the $k_v$-domain is equivalent to taking a thicker velocity slice in \cite{Esq03}. Rather thick slices were used in a numerical study by Padoan et al. (2006), which explains why they did not face the problems
related to shot-noise.

It is also important to test the VCS prediction of steepening of $P_1$ when the beam is made wide enough to emulate the low resolution regime. To do so we used steeper velocity spectrum, with slope $4$, which is less demanding in terms of the number of points along line of sight. If we meet condition (\ref{eq:Nz}) we have a good agreement with the predictions for both the high- and low-resolution regimes (see Fig. \ref{fig:VCS_4_00_2d}).

\section{Studies of turbulence with absorption lines from point sources} \label{sect:sal}

As clear from the earlier work (see LP00), as well as from our presentation above, the VCA and VCS are related techniques.
We must stress that although fields of applicability of discussed techniques intersect, they do not coincide. For instance, VCA requires less resolution over the velocity coordinate, because the range of scales can be obtained with spatial variations. Thus, potentially, it can be used for studies of gas with Mach number lower than what is required for VCS. 

\begin{figure}
\begin{center}
\plotone{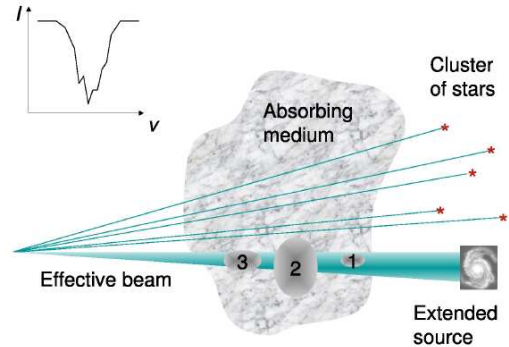}
\end{center}
\caption{VCS allows recovery of velocity statistics from absorption lines from stars. The whole $P_1$ is guaranteed to be in high-resolution mode, which provides better dynamical range over $k_v$. Numerical simulations show, that very few independent measurements are needed to gain required statistics (see Fig. \ref{fig:abs_10}). In the case of absorption against an extended source we have situation like shown on Fig. \ref{fig:VCS_res} for an emission line: the eddies can be in low (1), high (2) or intermediate (3) resolution mode. The effective beam is defined by the background object shape in this case, if it isn't resolved by the telescope. For absorption lines observed against CMB, we have the picture identical to shown on Fig. \ref{fig:VCS_res}\label{fig:abs}} 
\end{figure}

On the other hand, even if we have very poor statistics over angular coordinates and VCA does not't work, VCS still can be used. This opens new possibilities for studies of turbulence with absorption lines. In general, for absorption with extended sources both techniques can be readily applicable. As the spatial scale of the source decreases, the utility of the VCA decreases as well. Indeed, a lot of point-like sources are required to provide the spatial statistics required for the VCA. On the contrary, VCS utilizes sampling over different directions just to get different statistical
realizations of the turbulent process. Thus the number of point-like sources (e.g. stars) may be substantially reduced. The typical observational situation for probing turbulence with emission from point sources is illustrated in Fig. \ref{fig:abs}. What is the minimal number of sources of absorbed emission that are required for restoring the spectrum of turbulence between the sources and the observer? Below we answer this question using numerical simulations.


A general case of VCA studies with absorption lines is discussed in Lazarian \& Pogosyan (2008). There the criterium for restoring the turbulence signal from saturated absorption lines was obtained. For the sake of simplicity, here we discuss unsaturated absorption lines, arising from the medium with total optical depth less than unity. While, in general, a number of spectral lines from a single source may be used, we limit our study to the situation of a single absorption line. 
Our simulations show that only $~10$ independent measurements are needed to gain required statistics for $N_\sigma \approx 20$ (see Fig. \ref{fig:abs_10}). This opens avenues for studies of turbulence with background star even when the number of stars behind the turbulent volume is limited. No other presently known technique other than VCS can recover the velocity spectrum in this situation.

\begin{figure}
\begin{center}
\plotone{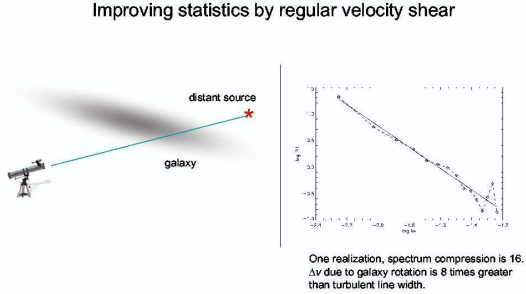}
\end{center}
\caption{VCS can be applied even to a single spectral line. Configuration can be as shown on this picture. Regular velocity shift from the galactic rotation results in statistics being sufficient to recover. $P_1$\label{fig:shift}} 
\end{figure}

We may note that the number of absorption sources may be even fewer than we mentioned above. This is the case, for instance, 
when the turbulent volume sampled along the line of sight contains subvolumes with different regular velocities. Such a velocity shift is possible due to galactic rotation, for instance. In this situation the limiting case for study is studying turbulence sampled by an absorption line from a point source along a single direction. In the presence of regular velocity difference between different turbulent volumes, the statistics may be rich enough to enable averaging of adjacent $k_v$-harmonics (see Fig. \ref{fig:shift}). 


Naturally, when one uses point-like sources to test turbulence with absorption lines, the effective resolution for the VCA is infinite. In the case of absorption lines arising from sources of finite angular size, the situation may be different. In particular, we may be also in the low resolution regime (see Fig \ref{fig:abs}).

\section{Discussion} \label{sect:disc}

Our study in the paper above clarifies a number of issues related to the studies of turbulence with the VCA and VCS techniques. In particular, our numerical testing proves the analytical expressions which are corner stones of the VCA and VCS techniques. We also identified the source of shot noise reported in the earlier numerical testing of the VCA, provided an analytical estimate of the velocity slice thickness required to eliminate the noise, and checked successfully that when the requirement of the velocity thickness is satisfied, the velocity spectrum can be successfully recovered.

In our treatment we stressed that the VCA and VCS techniques are closely related (cf. Eqs. (\ref{eq:P2}) and (\ref{eq:P1})) and therefore the presence of shot-noise in VCS inevitably affects the quality of VCA spectrum. We showed that the requirement given by Eq. (\ref{eq:Nz}) to eliminate the shot-noise in synthetic PPV cubes is rather demanding as the velocity spectrum gets less steep, which send a warning signal to brute-force attempts to test the techniques for an arbitrary index of turbulence spectrum. We confirm, however, an earlier claim in Esquivel et al. (2003) that the shot-noise interference is, exclusively, a problem of synthetic observations, rather than the real ones. Shot-noise will appear in real observations too, when the number of received photons is low enough. Incidentally, the condition (\ref{eq:Nz}) gives in this case the required number of photons and can be used for the estimation of observational time required. 

The limitation of our present study was that we dealt only with testing of the asymptotical solutions for the VCA and VCS techniques. Real data with its limited dynamical range of observed velocity fluctuations may benefit from fitting the observations of the integral expressions rather than the asymptotical solutions. Indeed, the actual spectrum in the velocity slice may be thin for large eddies and thick for small eddies. Therefore asymptotical solutions may fit only part of the actual spectrum of the slice, which will deviate from a power-law. We shall deal with the numerical testings of these situations elsewhere.

\acknowledgements
The authors thank Dmitry Pogosyan for his input. AC and AL acknowledge the support from the Center for Magnetic Self-Organization in Laboratory and Astrophysical Plasmas and NSF grant AST 0808118. We also thank our referee Anthony Minter for his valuable input.


\appendix

\section{Spectral Line} \label{sect:sig}

Let us write out the space-velocity distribution of density:
\begin{equation} \label{eq:dens}
n_0(\vc{r},v)
  = n_0(\vc{r}) \varphi(v-v_{macro}(\vc{r})),
\end{equation}
\noindent where $\varphi(v)$ is Maxwellian distribution of the emitting atoms (ions):
\begin{equation} \label{eq:maxw}
\varphi(v)
  = \frac{1}{\sqrt{2\pi\beta}} \exp\left(-\frac{v^2}{2\beta}\right), \; \beta = \frac{k_B T}{m}
\end{equation}
\noindent $k_B$ is Boltzmann's constant and $n_0(\vc{r})$ is an integral density. 

In these terms emissivity can be written as follows:
\begin{equation} \label{eq:emiss}
\eps_0(\vc{r},v)
  = \eps_0(\vc{r}) \varphi(v-v_{macro}(\vc{r})),
\end{equation}
\noindent where
\begin{equation} \label{eq:emiss_lin}
\eps_0(\vc{r})
  = \gamma_1 n_0(\vc{r})
\end{equation}
-- for linear emissivity law, or
\begin{equation} \label{eq:emiss_quad}
\eps_0(\vc{r})
  = \gamma_2 n_0^2(\vc{r})
\end{equation}
-- for quadratic emissivity law,

Here $\gamma_1$, $\gamma_2$ are the coefficients depending on the correspondent cross-sections, and $v$ defines emission frequency shift.

Let us write out the elementary power\footnote{we neglect the self-absorption effects} , radiated by the elementary volume $\D{\vc{r}}\D{v}$ into the solid angle $\D{\Omega}$:
\begin{equation} 
\begin{array}{ll}
\D{P} 
  & = \eps_0(\vc{r},v) \frac{\D{\Omega}}{4\pi} \D{\vc{r}} \D{v} \\
  & = \frac{1}{4\pi r^2} \eps_0(\vc{r}) \varphi(v-v_{flow}) \D{s} \D{\vc{r}} \D{v}
\end{array}
\end{equation}
\noindent where $\D{s}$ is an elementary area at the distance $r$ from the source.

Then we can express the elementary signal at the spectrometer output as follows:
\begin{equation} 
\D{S} 
  = \frac{1}{2k_B} \frac{A_{eff} \D{P}}{\D{s}} \frac{c}{f_0} f_s(v-v_0) w_{b,1}(\theta,\phi) \D{\vc{r}} \D{v}
\end{equation}
\noindent where $A_{eff}$ is effective area of the instrument, $f_s(v)$ is the amplitude-frequency response of the spectrometer normalized to the integral value, $f_0$ and $v_0$ are central frequency and velocity of the channel and $w_{b,1}(\theta,\phi)$ is the instrument beam normalized to the maximal value. $S$ is measured in Kelvins.

Accounting for
\begin{equation} 
A_{eff} 
  = \eps_a \frac{\lambda_0}{\Omega_a}
\end{equation}
\begin{equation} 
\Omega_a 
  = \int w_{b,1}(\theta,\phi) \D{\Omega}
\end{equation}
\noindent where $\Omega_a$ is the beam solid angle, $\lambda_0$ is central wavelength and $\eps_a$ is the aperture efficiency and defining
\begin{equation} 
w_b 
  \equiv \frac{w_{b,1}}{\Omega_a}
\end{equation}
\noindent we obtain:
\begin{equation} 
\D{S} 
  = \eps_a \frac{\lambda_0^3}{2k_B} \frac{1}{4\pi r^2} \varphi(v-v_{flow}(\vc{r})) f_s(v-v_0) w_b(\vch{e},\vch{r}) \D{\vc{r}} \D{v}
\end{equation}
where $\vch{e}$ is the beam direction. Integrating $\D{S}$ we finally have the emission line profile:
\begin{equation} 
S(\vch{e},v_0)
  = \frac{\eps_a \lambda_0^3}{8\pi k_B} \int  \D{\vc{r}} \frac{w_b(\vch{e},\vch{r})}{r^2} \eps_0(\vc{r}) f(v_0-v_{flow}(\vc{r}))
\end{equation}
where effective channel sensitivity function $f$ is defined as follows:
\begin{equation} \label{eq:effsens}
f(v) 
  \equiv \int_{-\infty}^\infty \varphi(v'+v) f_s(v') \D{v'}
\end{equation}

Random field $\eps_0(\vc{r})$ is not homogeneous, at least because the emitting structure is limited in space. To model this we introduce a homogeneous field $\eps(\vc{r})$ and a deterministic factor $w_\eps(\vc{r})$ setting up the borders of an observed object. We will also ``pack'' into $\eps(\vc{r})$ all constant factors:
\begin{equation} \label{eq:eps0}
w_\eps(\vc{r}) \eps(\vc{r})
  \equiv \frac{\lambda^3}{8\pi k_B} \eps_0(\vc{r}).
\end{equation}
Introducing a window function $w$ as follows\footnote{This composite window function is a product of the 3D instrument sensitivity function and the window function defining the object extent.}:
\begin{equation}
w(\vch{e},\vc{r})
  \equiv \frac{1}{r^2} w_b(\vch{e},\vch{r}) w_\eps(\vc{r}),
\end{equation}
we finally obtain:
\begin{equation} 
S(\vch{e},v_0)
  = \int w(\vch{e},\vc{r}) \D{\vc{r}} \eps(\vc{r}) f(v_r(\vc{r}) + v_r^{reg}(\vc{r}) - v_0).
\end{equation}

\section{Code}

The simulation programs are written in C++ in single-processor and MPI variants. The source of simulation programs is available at \sourcedir, see Tab. \ref{tab:progs} for the complete list.


\begin{deluxetable}{llll}
\tablecaption{List of simulation programs\label{tab:progs}}
\tablehead{
  \colhead{program} & 
  \colhead{dimensions} & 
  \colhead{techniques} & 
  \colhead{architecture}
}
\startdata
spect1d    & 1 & VCS & single processor\\ 
spect2d    & 2 & VCS & single processor\\ 
spect3d    & 3 & VCS & single processor\\ 
p\_spect2d & 2 & VCS & MPI\\ 
p\_spect3d & 3 & VCS & MPI\\ 
vca3d      & 3 & VCA & single processor\\ 
p\_vca2d   & 2 & VCA & MPI\\ 
p\_vca3d   & 3 & VCA & MPI\\
\enddata
\tablecomments{The programs are available at \protect{\sourcedir}}
\end{deluxetable}


\clearpage

\begin{figure}
\begin{center}
\begin{tabular}{lll}
\inctab{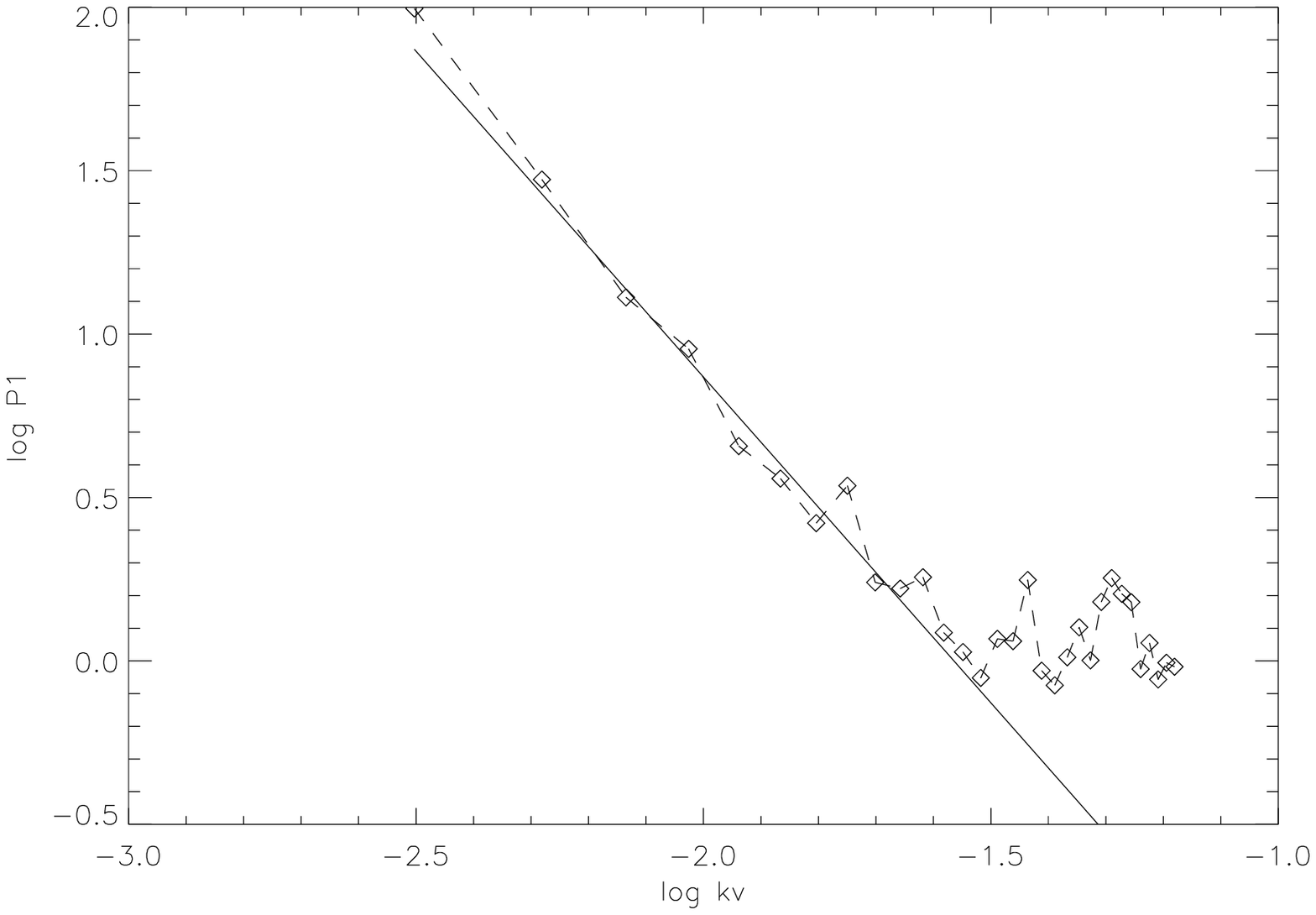} &
\inctab{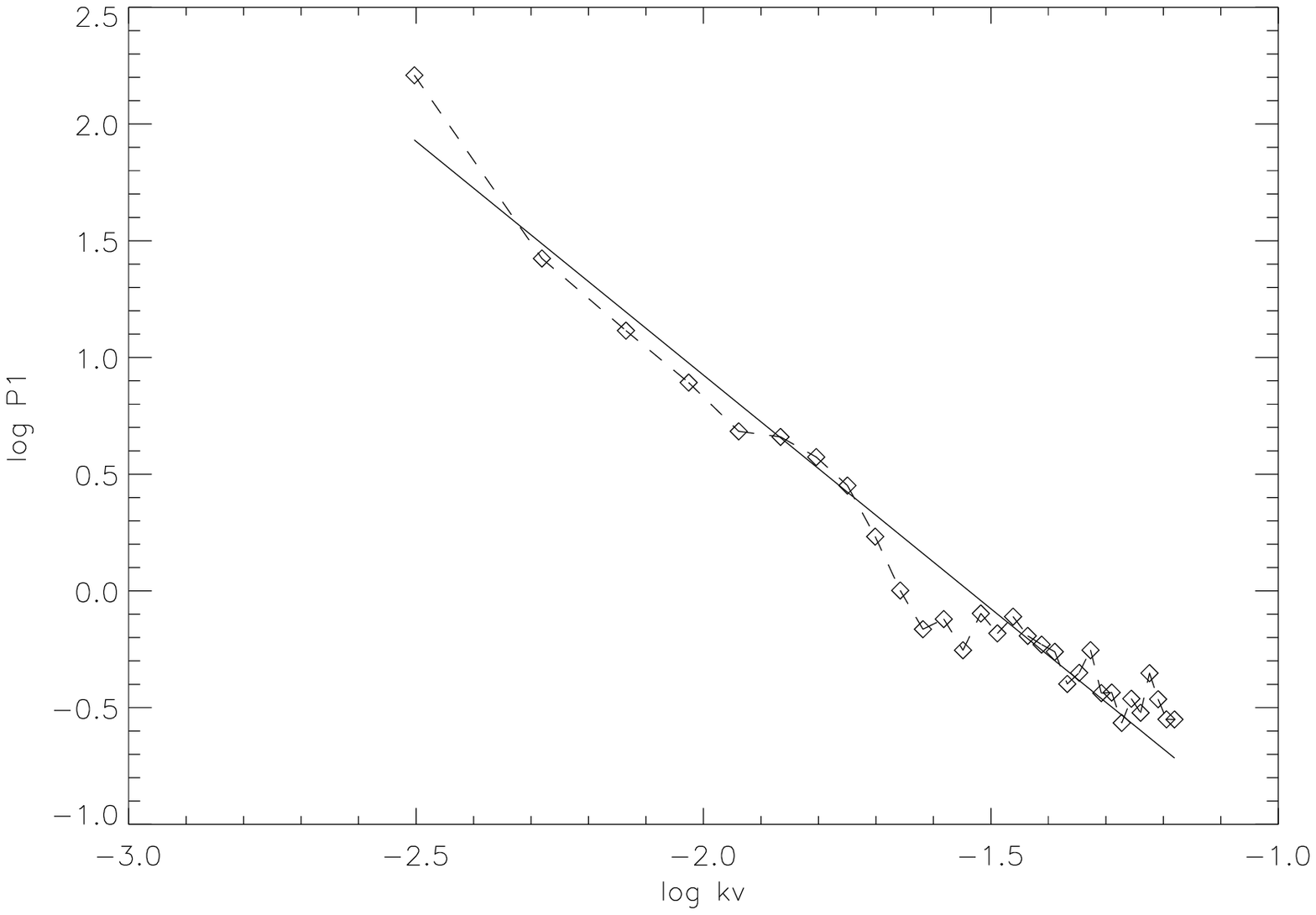} &
\inctab{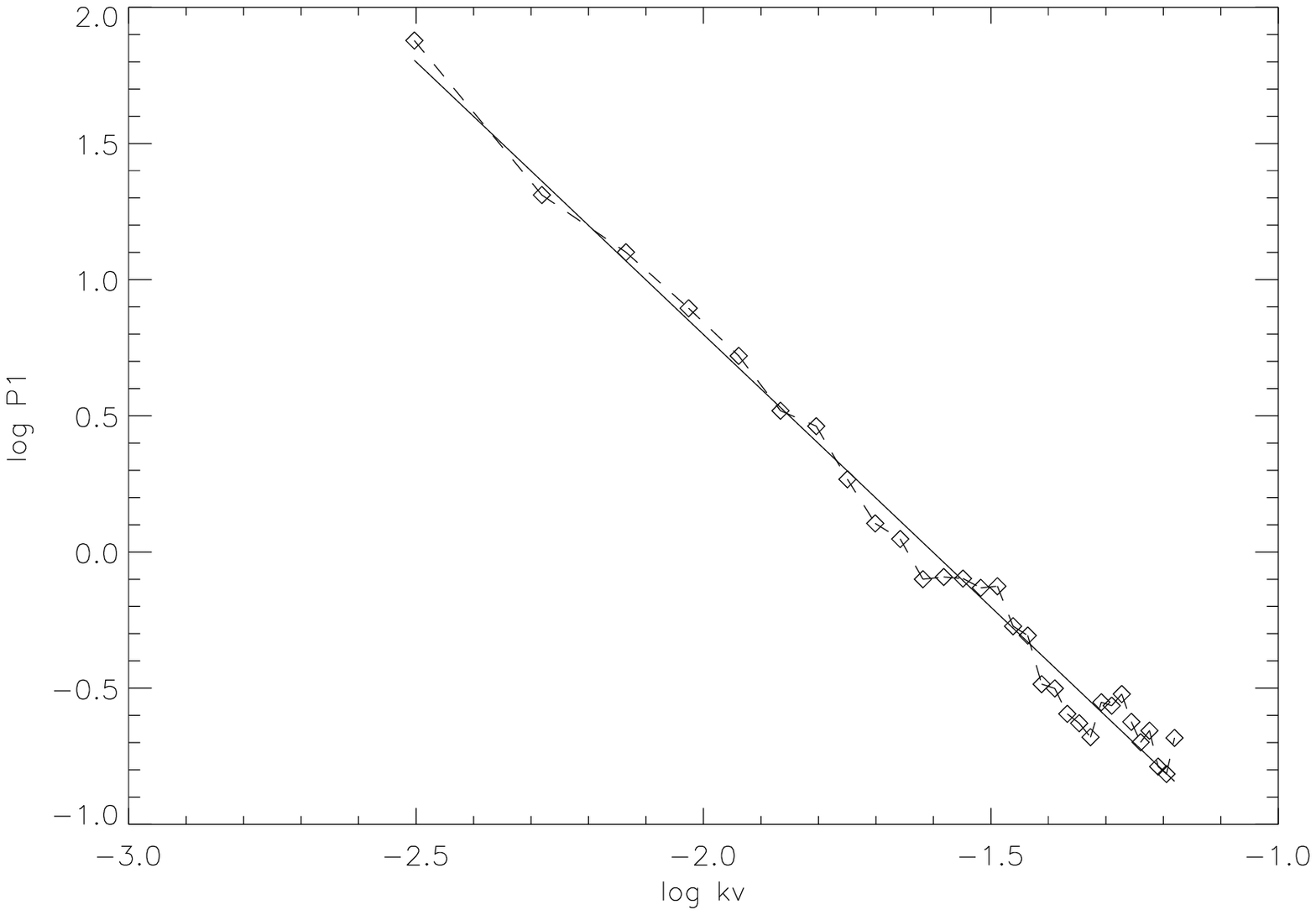} \\
\inctab{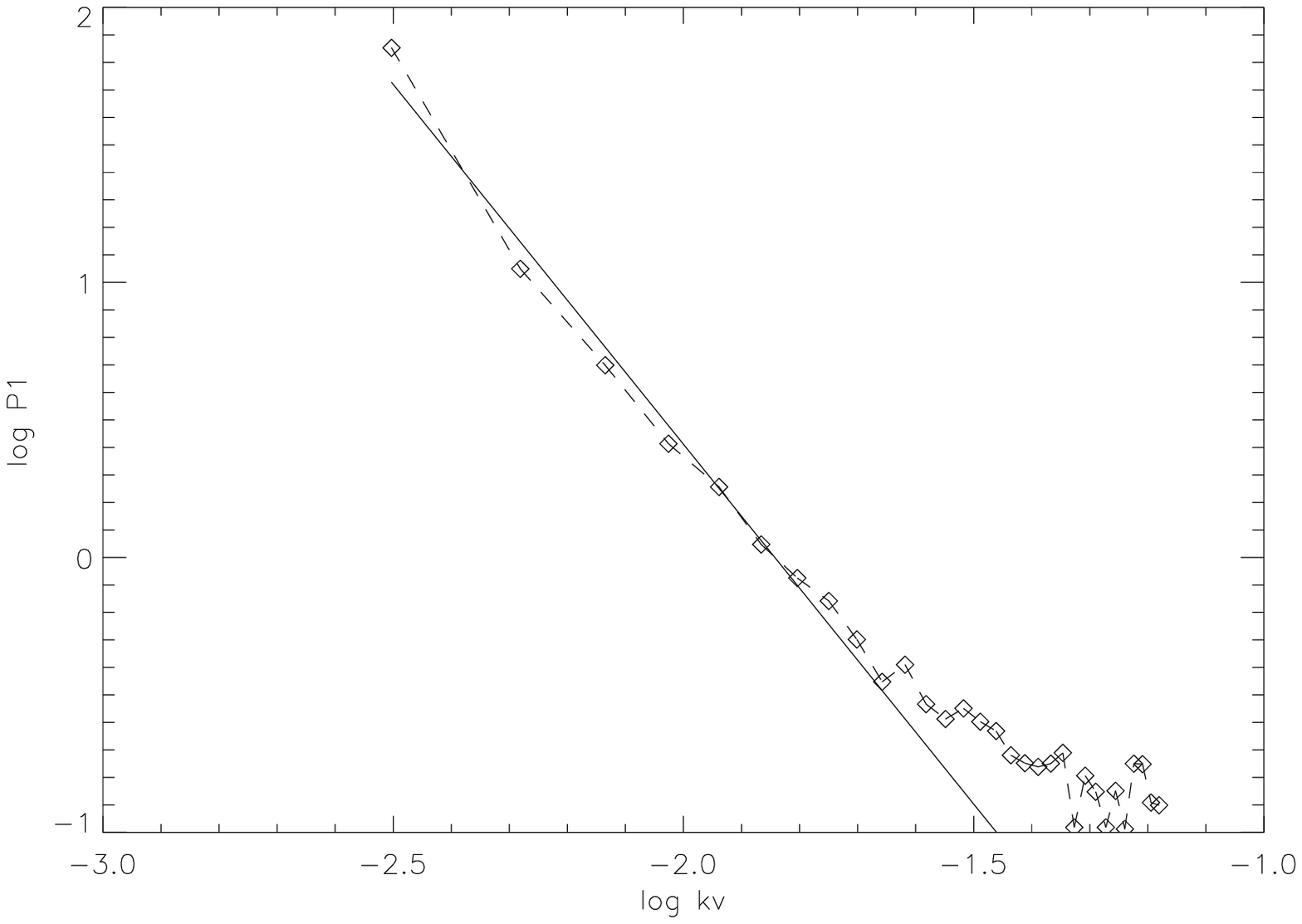} &
\inctab{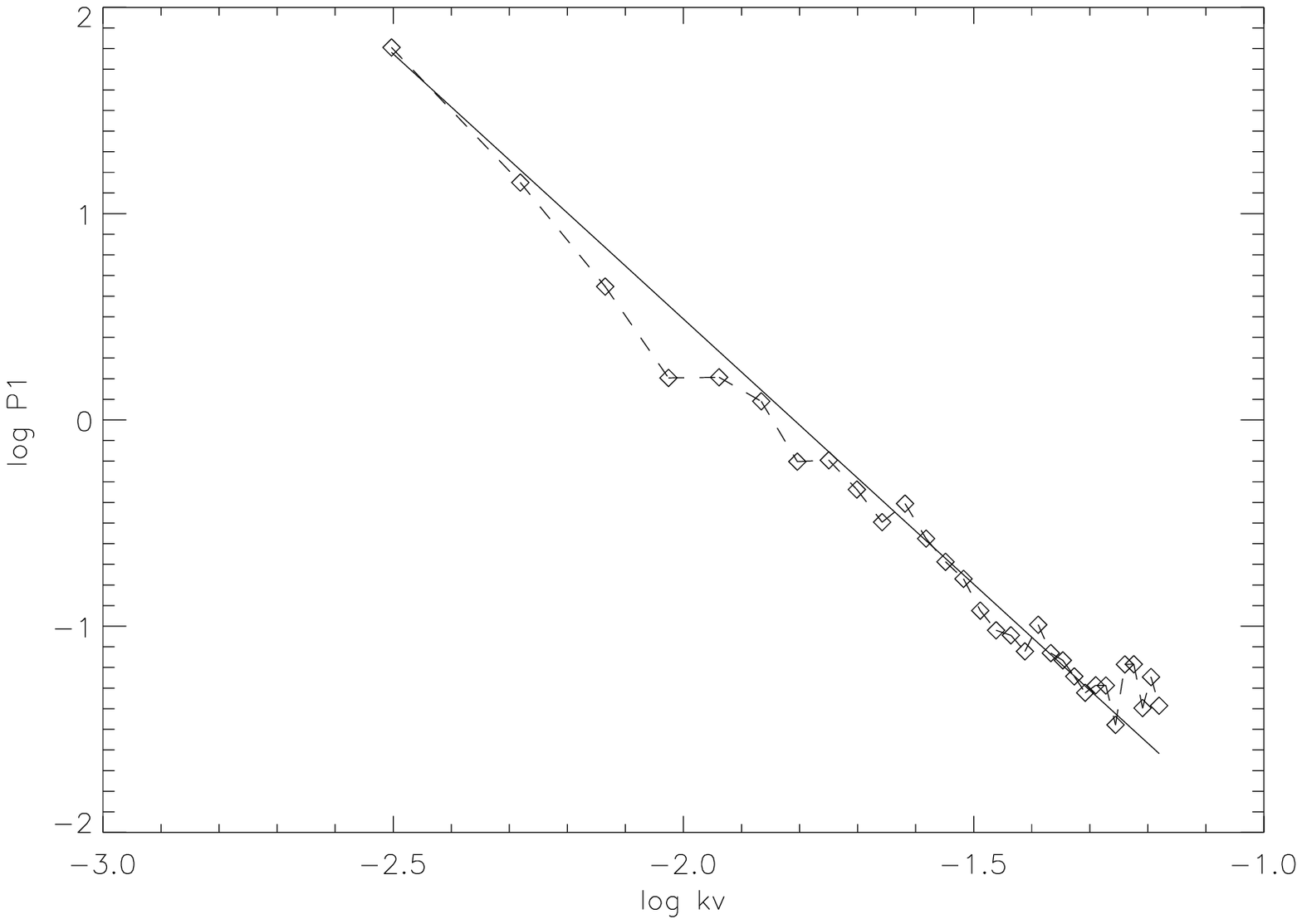} &
\inctab{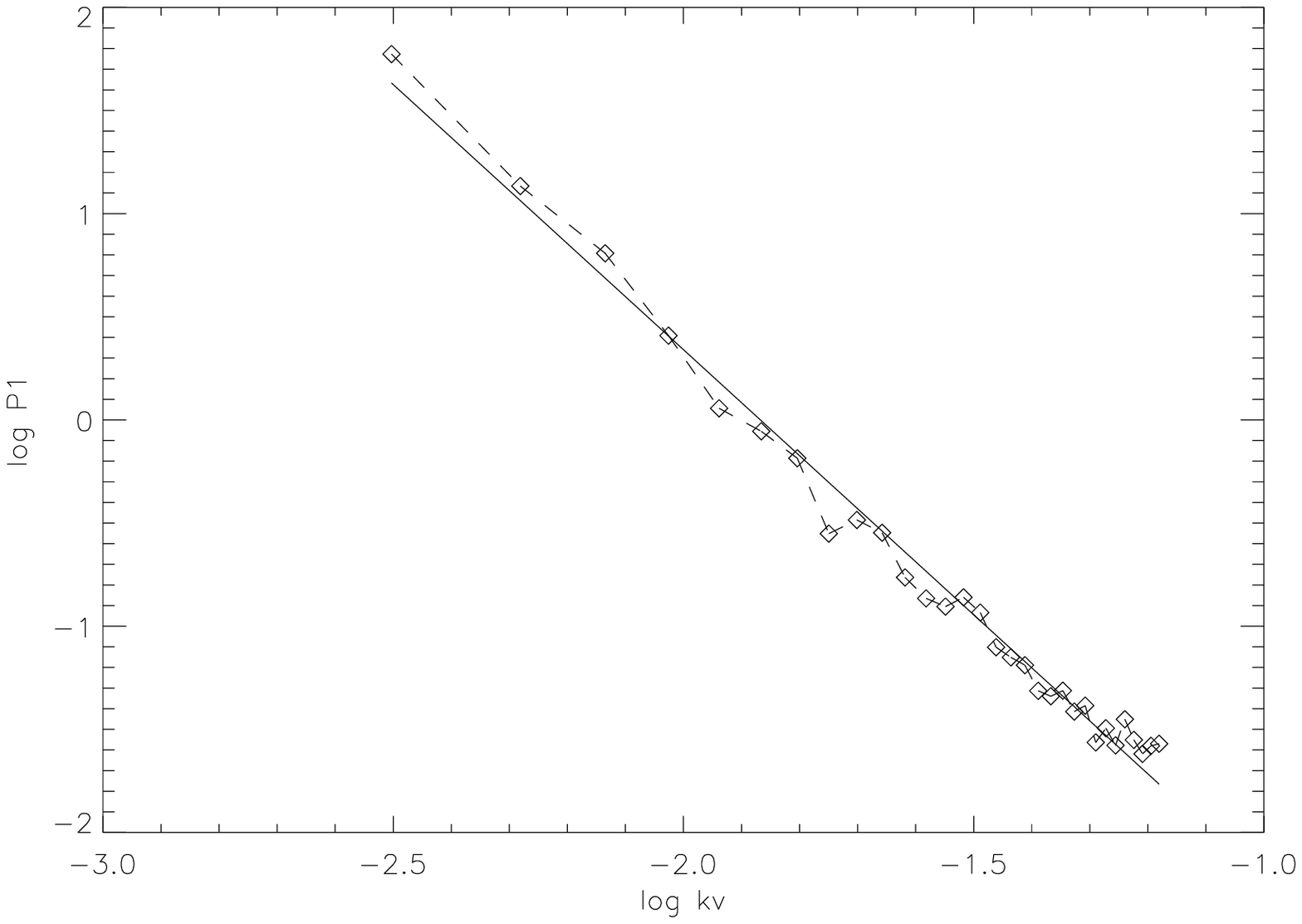} \\
\inctab{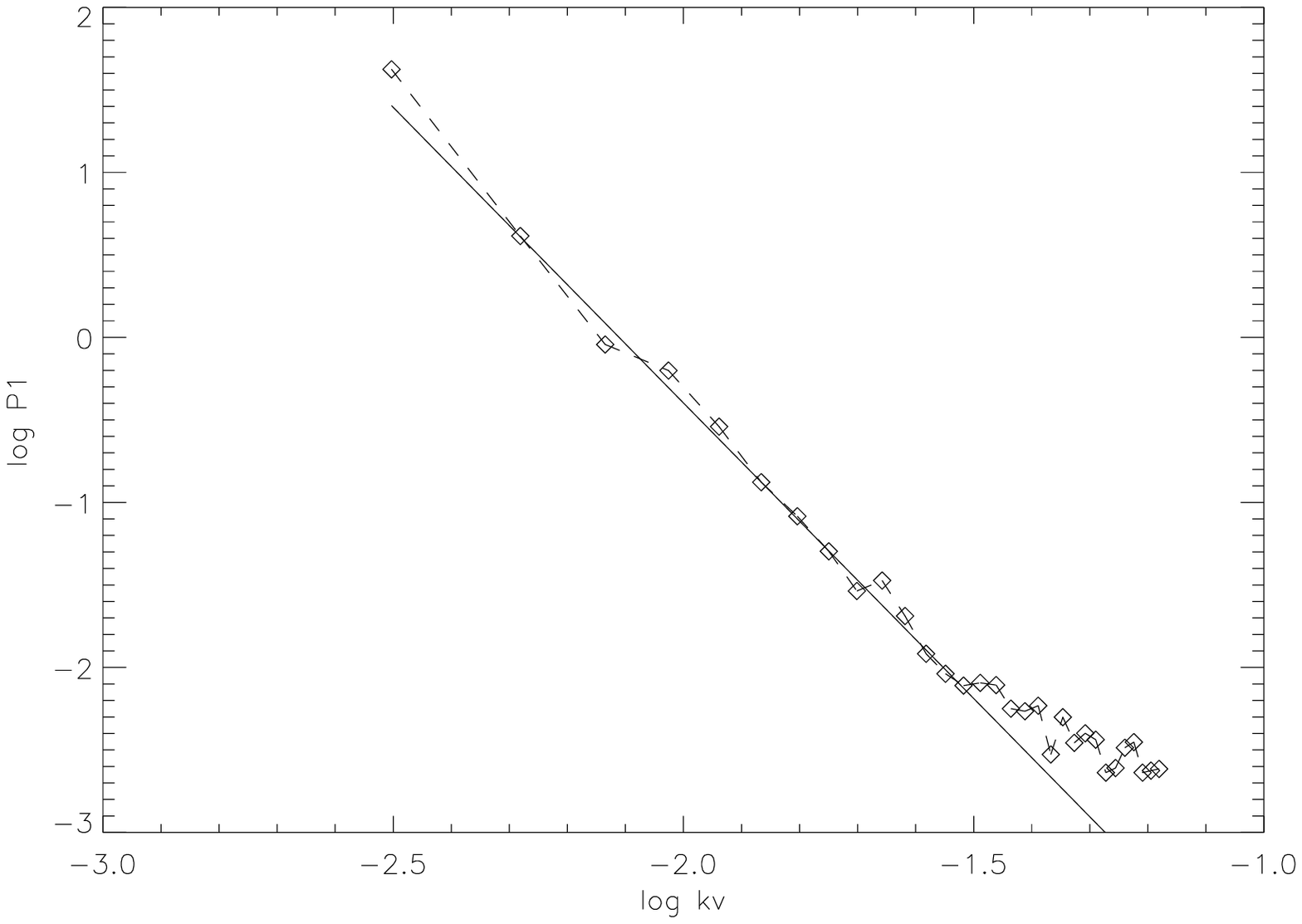} &
\inctab{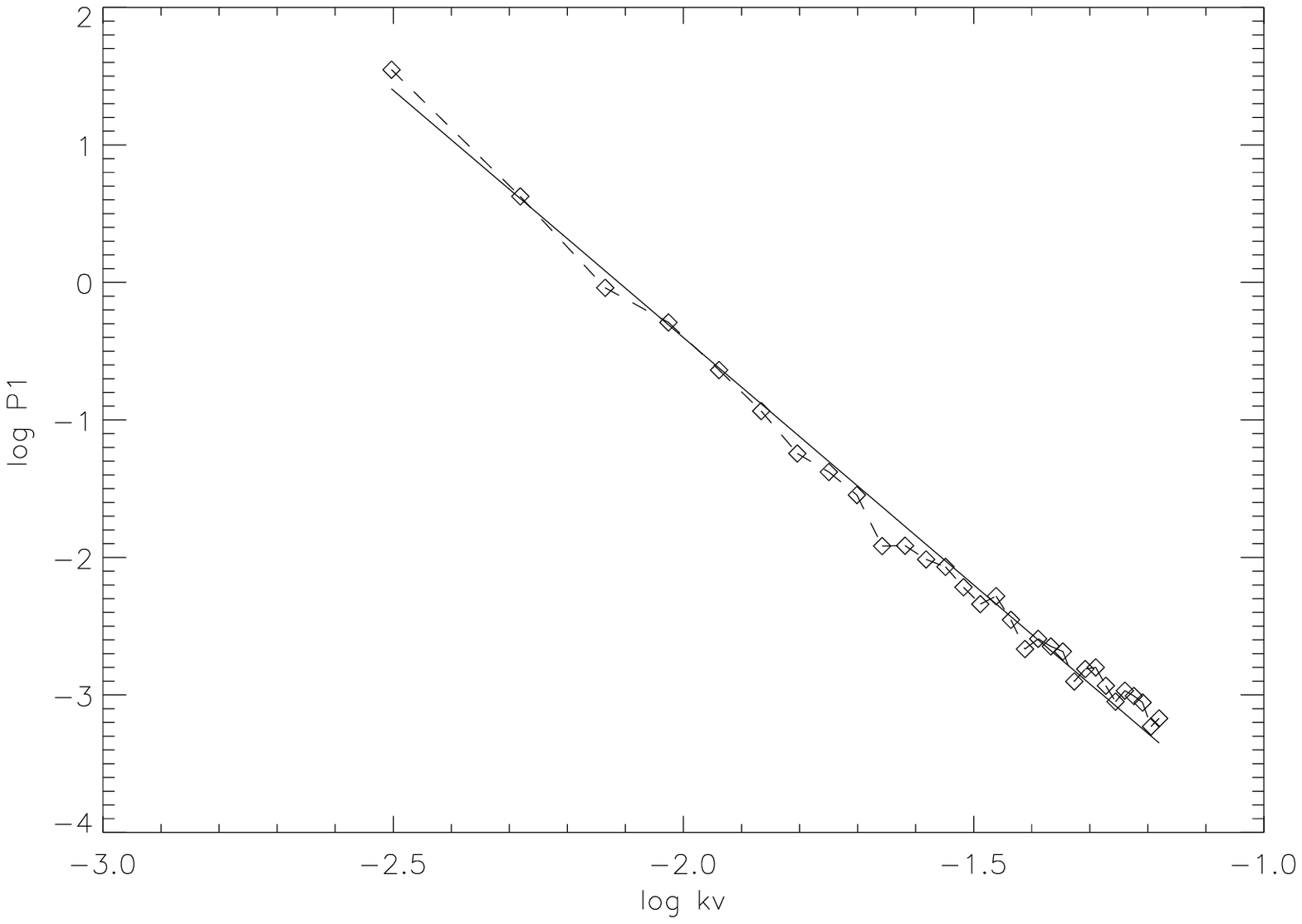} &
\inctab{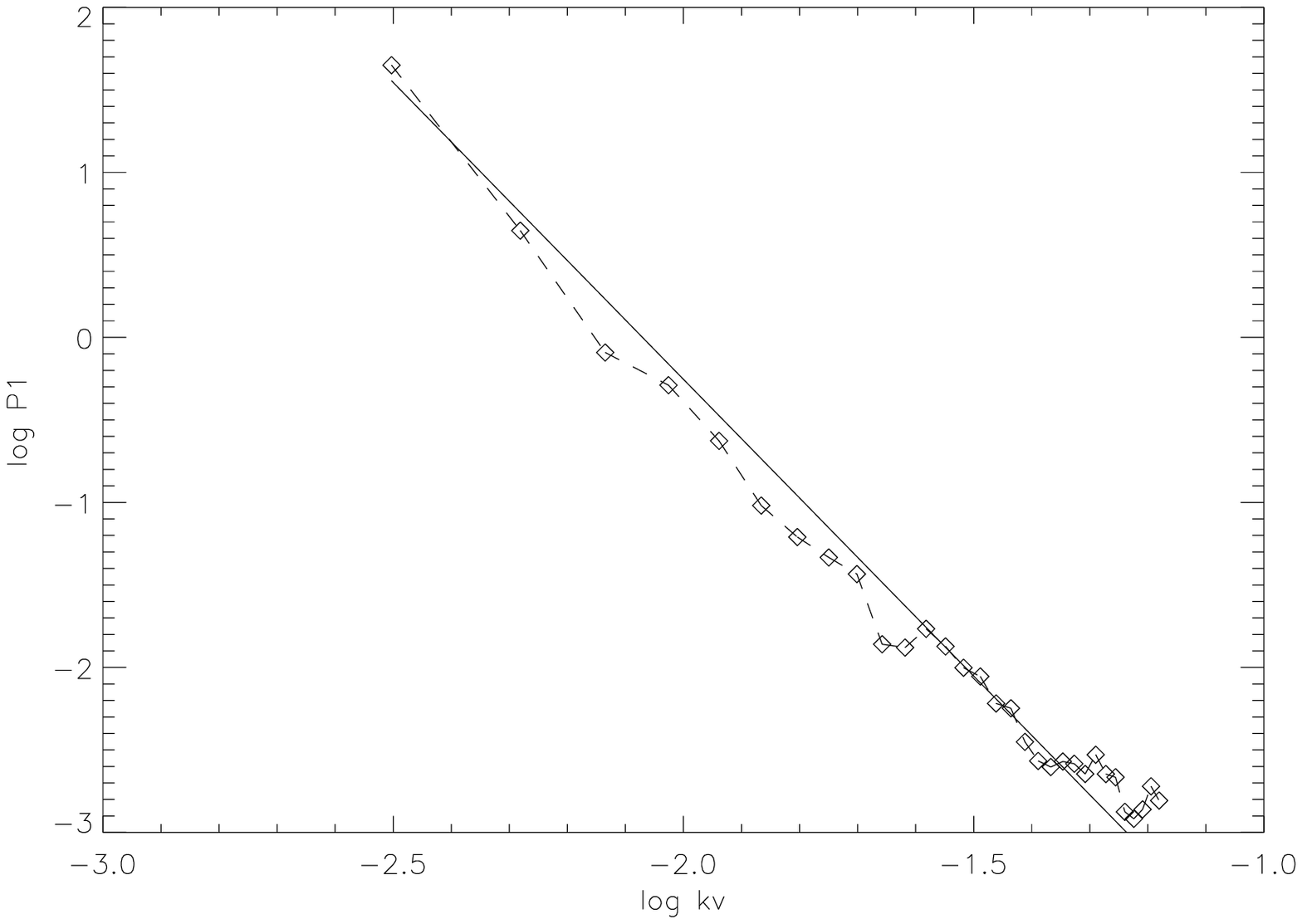}
\end{tabular}
\end{center}
\caption{Illustration to the requirement [\ref{eq:Nz}] to the number of points over line of sight. The rows correspond to different velocity spectral index $\alpha_v$ (4, 3.78 and 3.56 from top to bottom). VCS spectra in the middle column correspond to $N_z$ close to the estimation ($2^{14}$, $2^{17}$ and $2^{23}$ respectively), in the left column $N_z$ is 4 times lower, in the right -- 2 times greater. The cases with low $N_z$ exhibit shot-noise, which results in points getting above the solid line, which represents the analytical expectation.\label{fig:Nz_req}} 
\end{figure}

\begin{figure}
\begin{center}
\plottwo{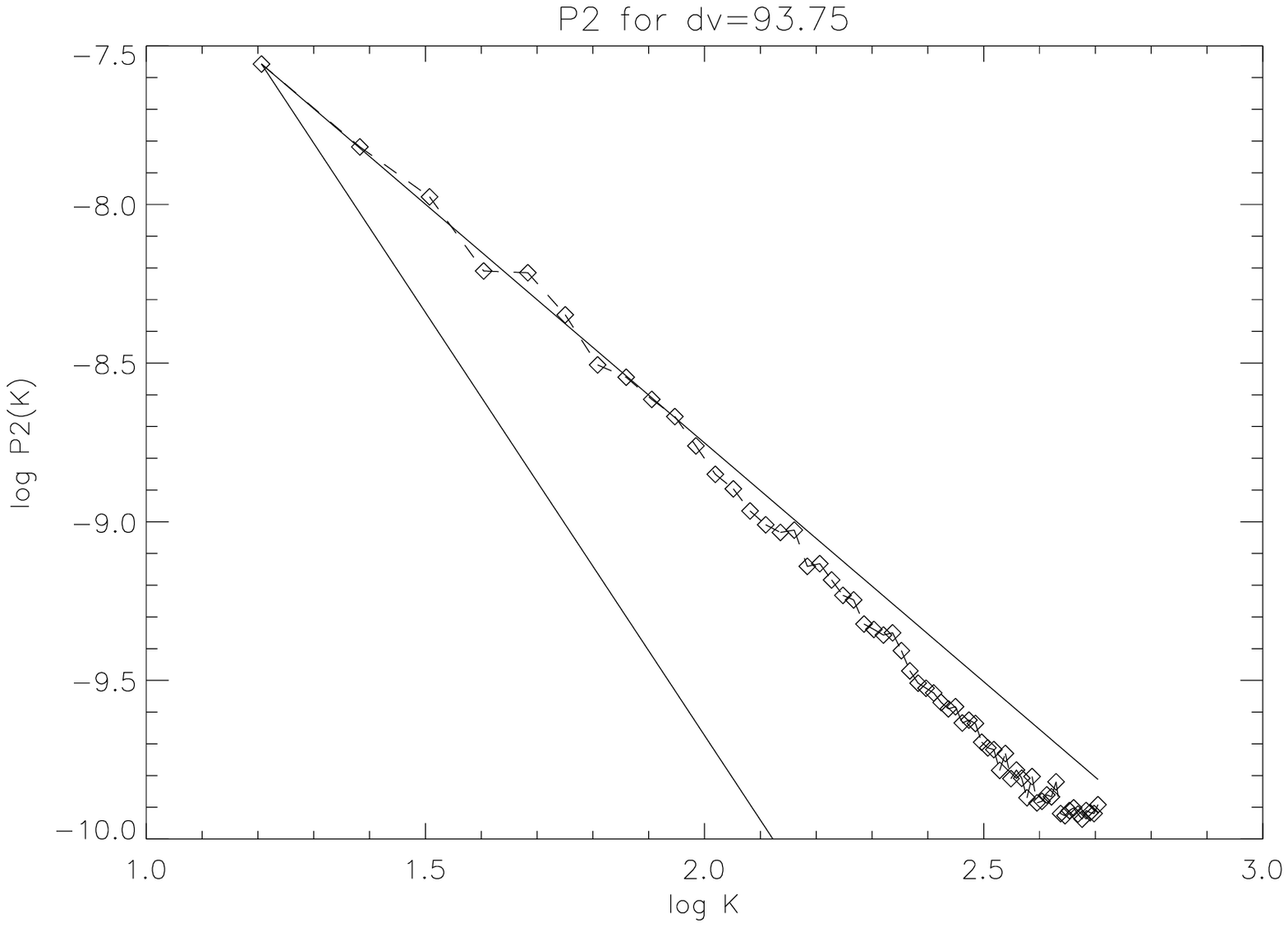}{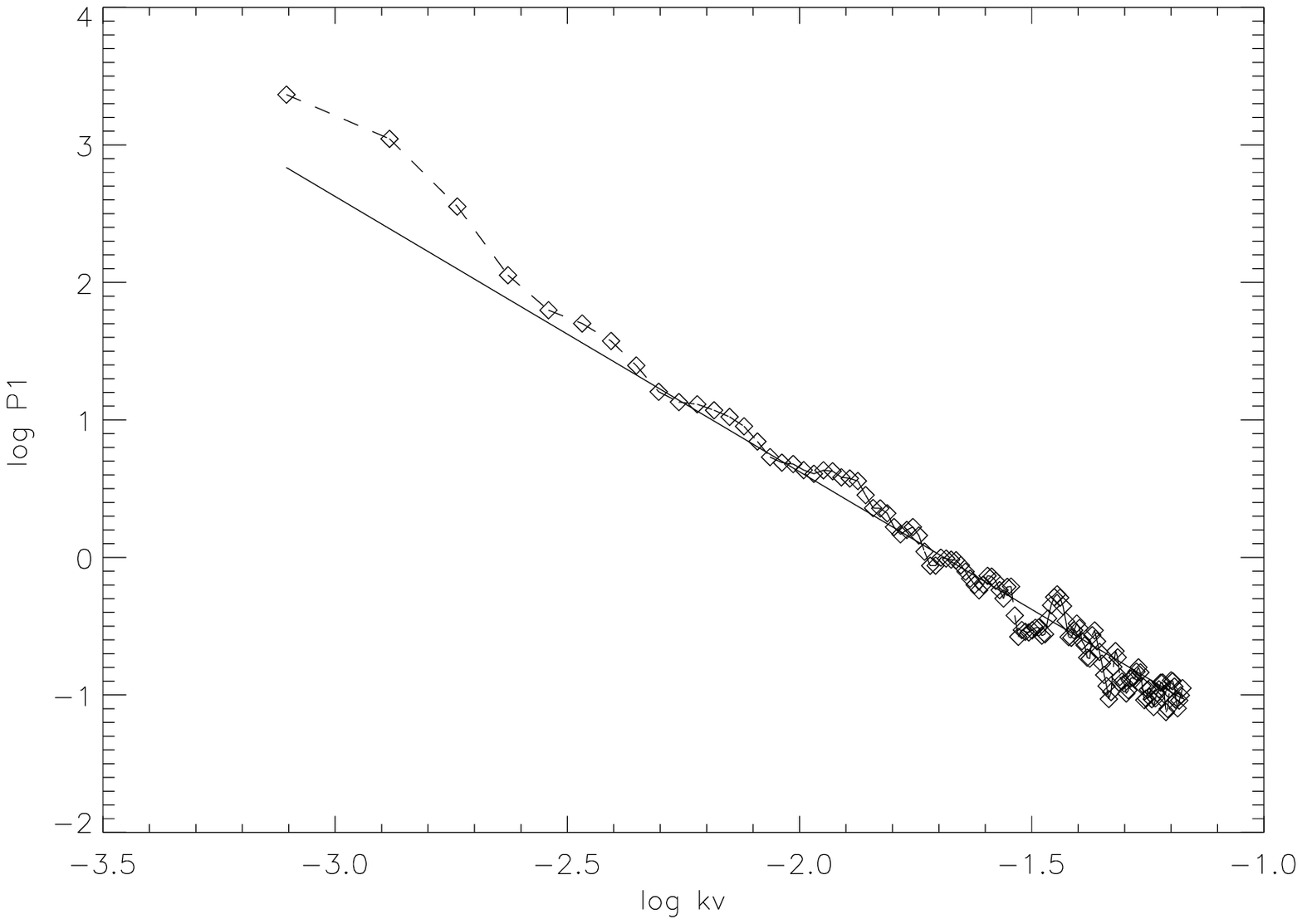}
\end{center}
\caption{Simulations with the number of points over line of sight $N_z=32768$, sufficient to suppress shot-noise ($N_z=20000$ is required). Left: VCA spatial spectrum (2D simulation), shallower solid line -- expected velocity-dominated spectrum, steeper solid line -- density-dominated spectrum. Right: correspondent VCS spectrum for high resolution mode (1D simulation), solid line -- expected slope. Velocity spectral index is $4$. \label{fig:VCA_VCS_4_00}}
\end{figure}

\begin{figure}
\begin{center}
\plottwo{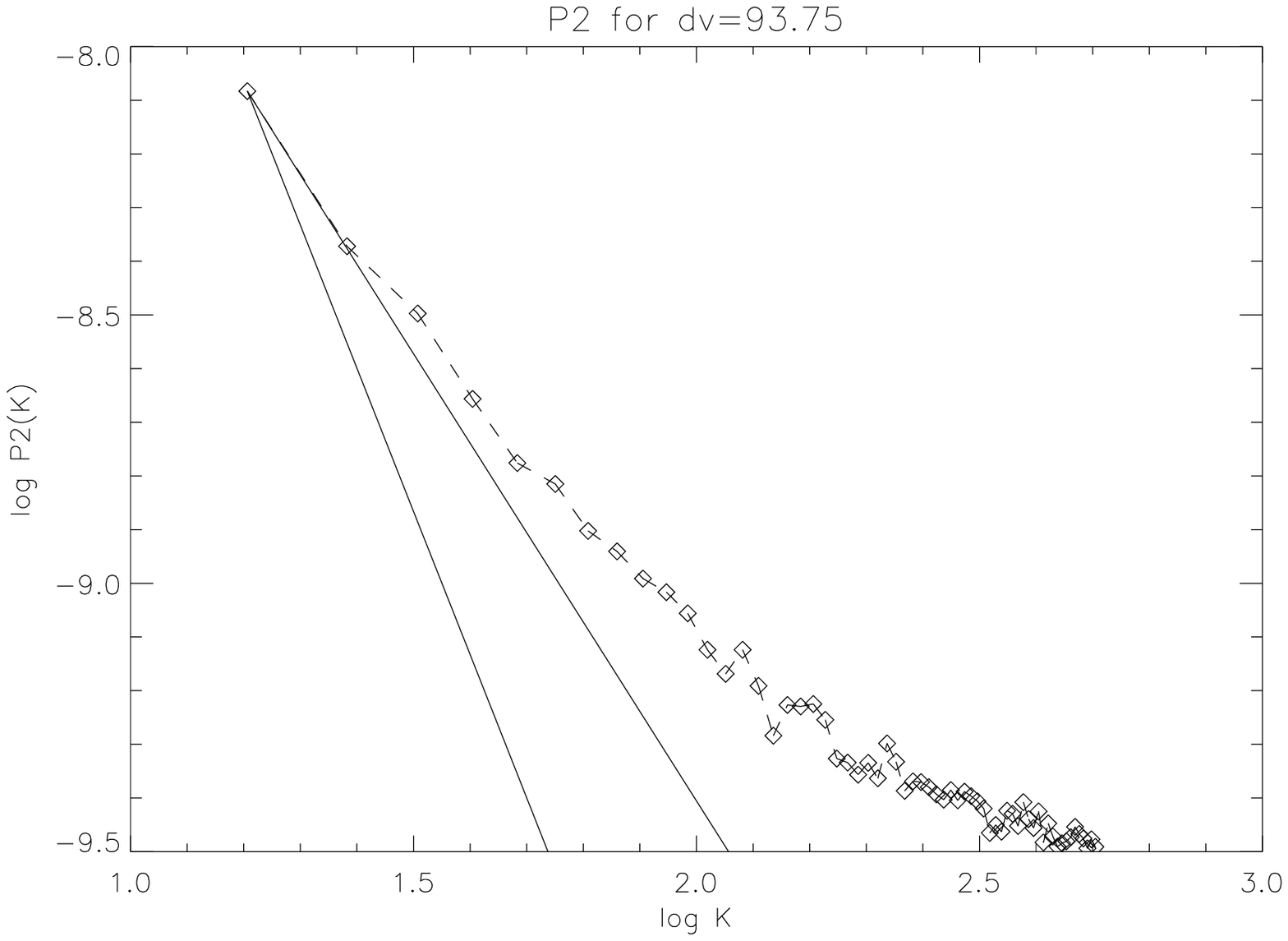}{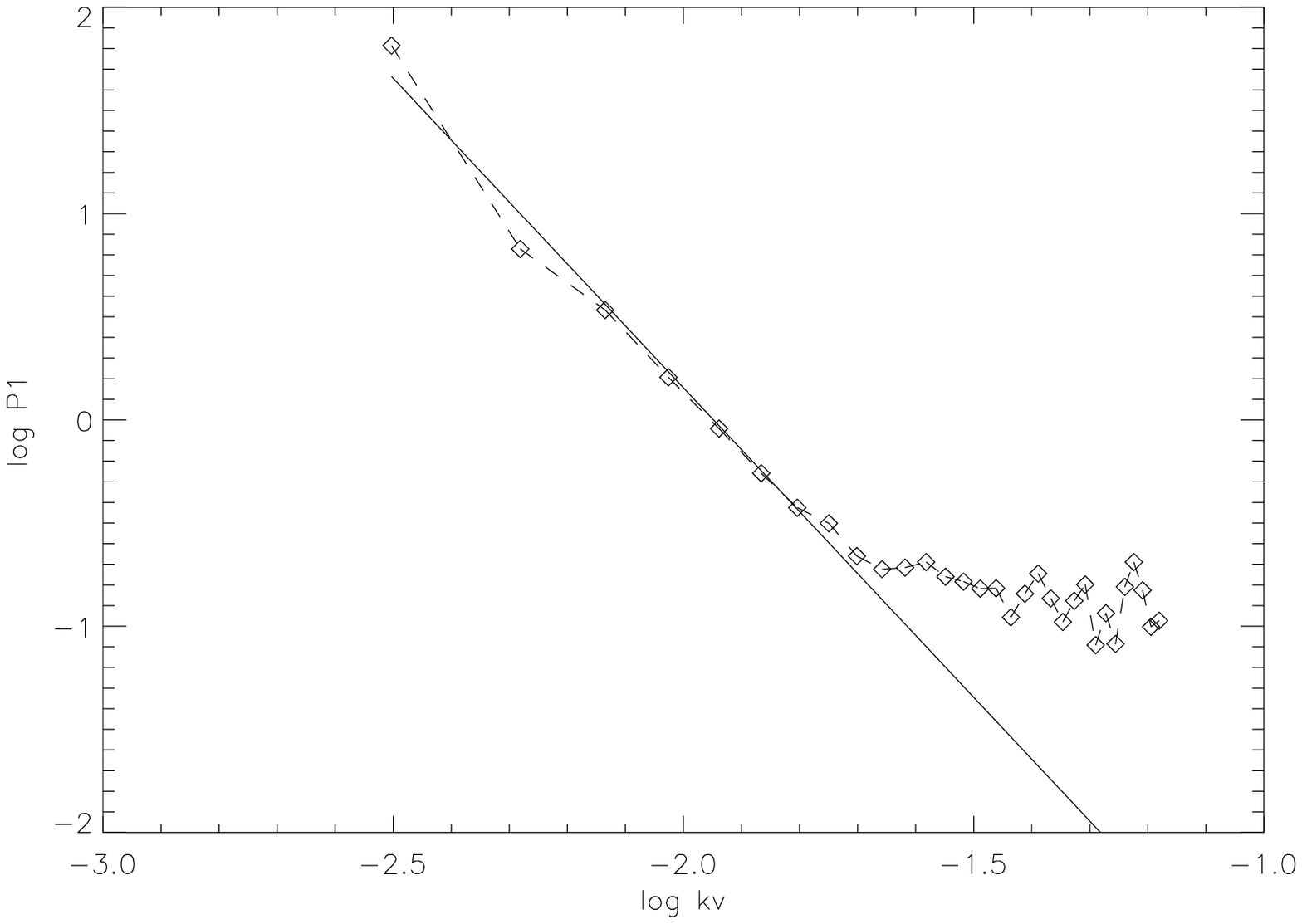}
\end{center}
\caption{Simulations with insufficient number of points over the line of sight ($N_z=32768$ instead of $N_z \approx 420000$). Left: VCA spatial spectrum (2D simulation), shallower solid line -- expected velocity-dominated spectrum, steeper solid line -- expected density-dominated spectrum; the spectrum is distorted (the velocity-dominated one is expected). Right: correspondent VCS spectrum for high resolution mode (1D simulation), solid line -- expected slope, shot-noise clearly visible. Velocity spectral index is $11/3$, which implies larger minimal $N_z$, than for the case on Fig. \ref{fig:VCA_VCS_4_00}.   \label{fig:VCA_VCS_3_67_unfilt}} 
\end{figure}

\begin{figure}
\begin{center}
\plottwo{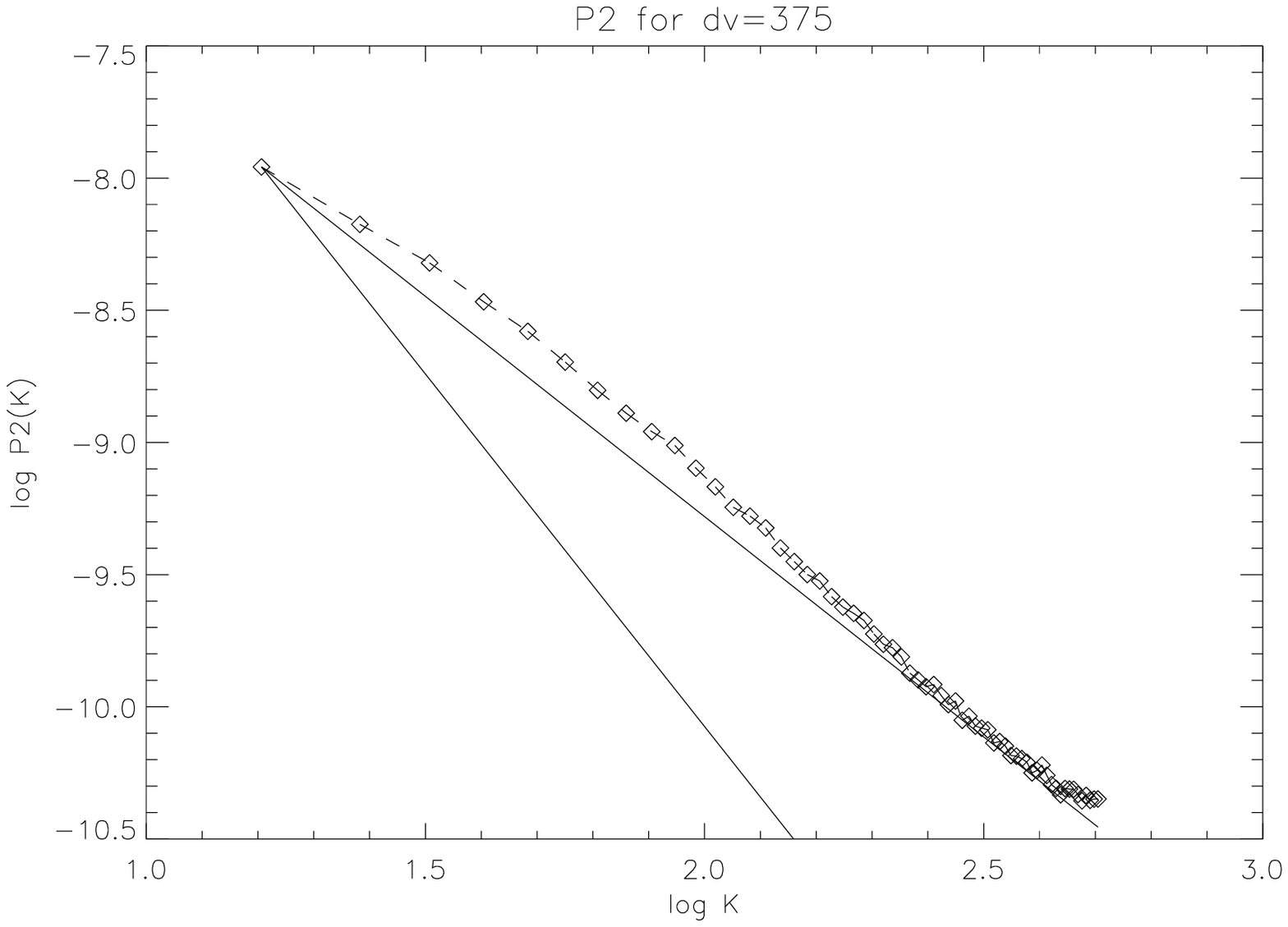}{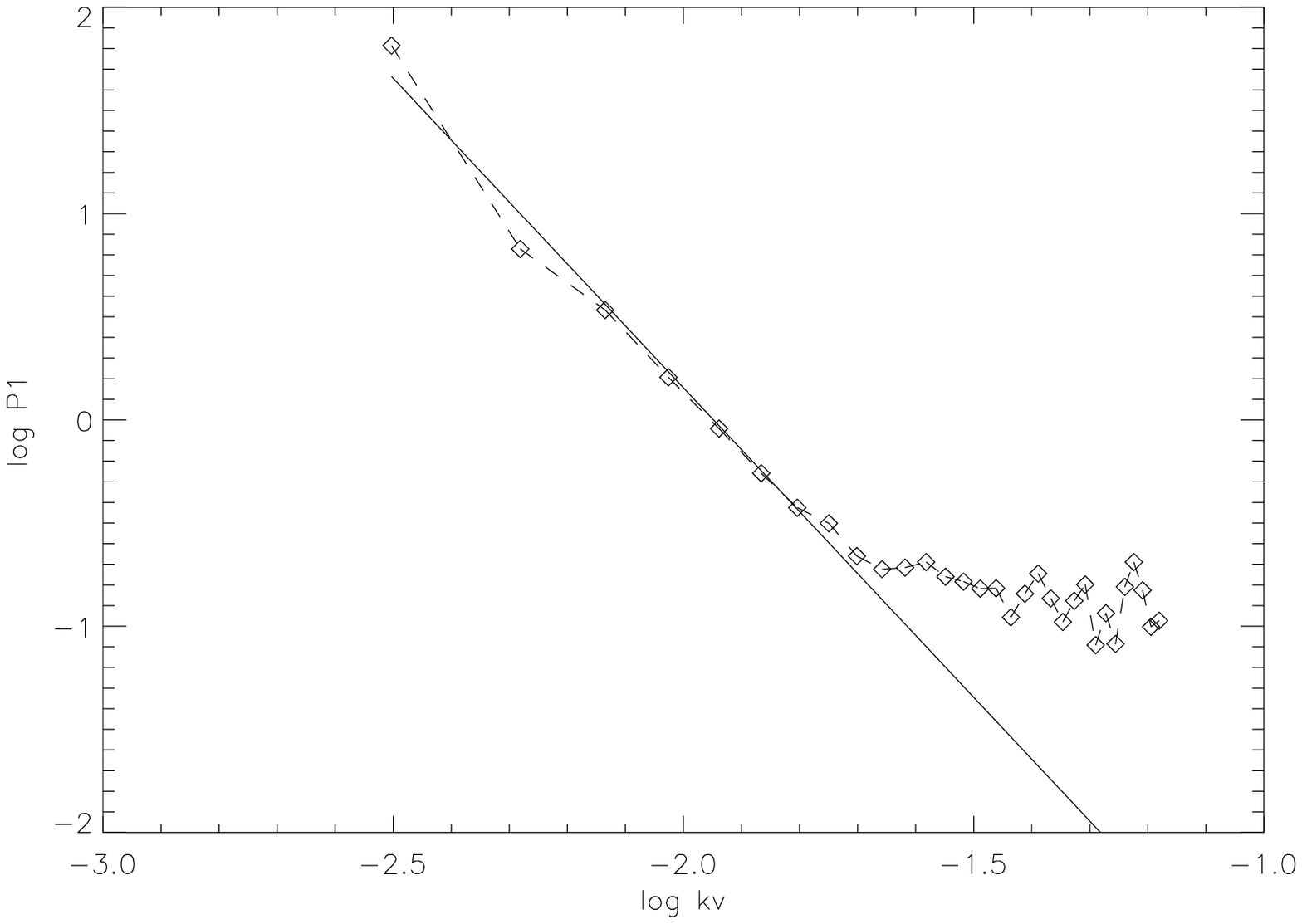}
\end{center}
\caption{The same as Fig. \ref{fig:VCA_VCS_3_67_unfilt}, but with noisy $k_v$-harmonics filtered out when calculating the VCA spectrum (left). \label{fig:VCA_VCS_3_67_filt}} 
\end{figure}

\begin{figure}
\begin{center}
\plottwo{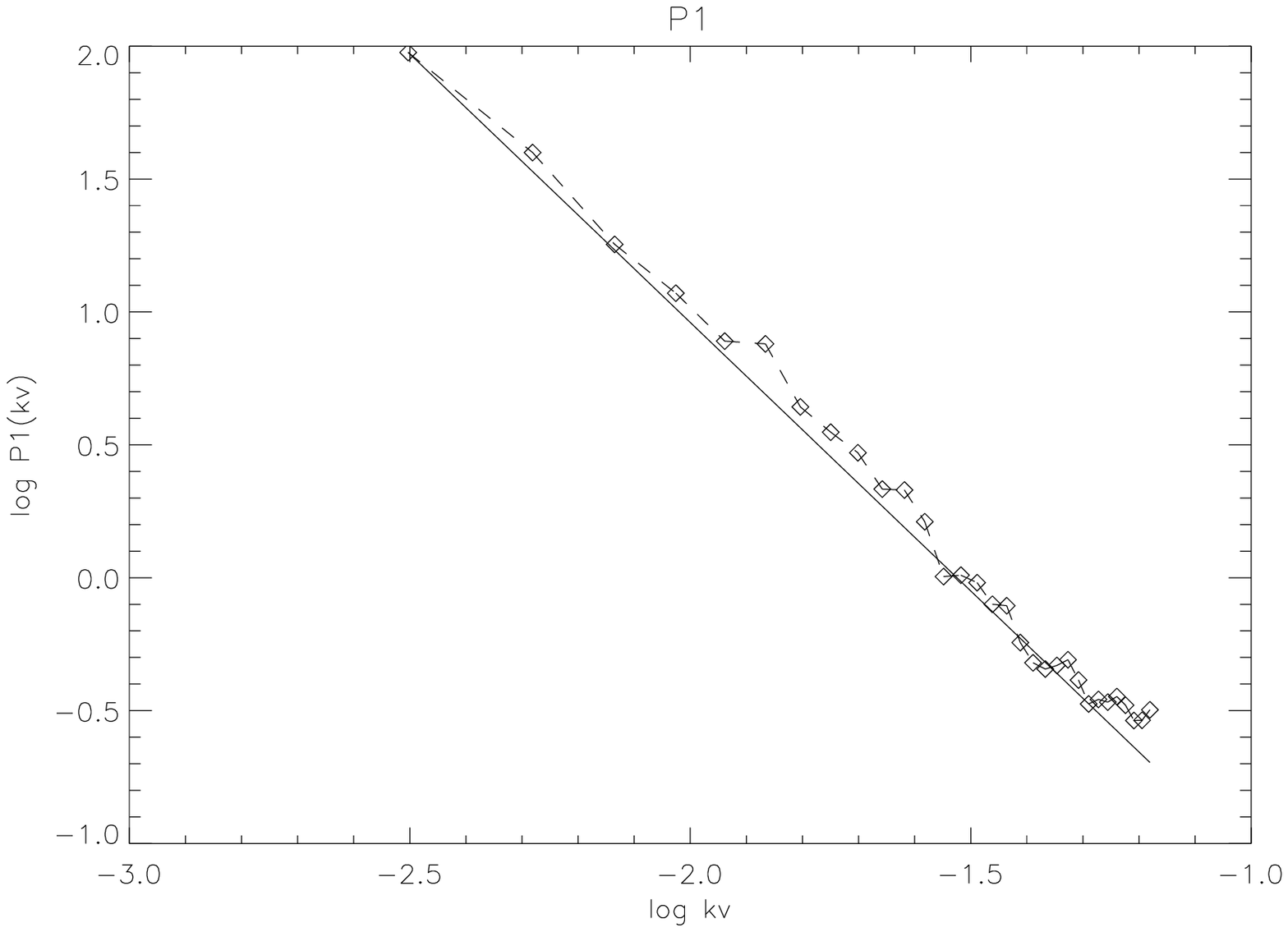}{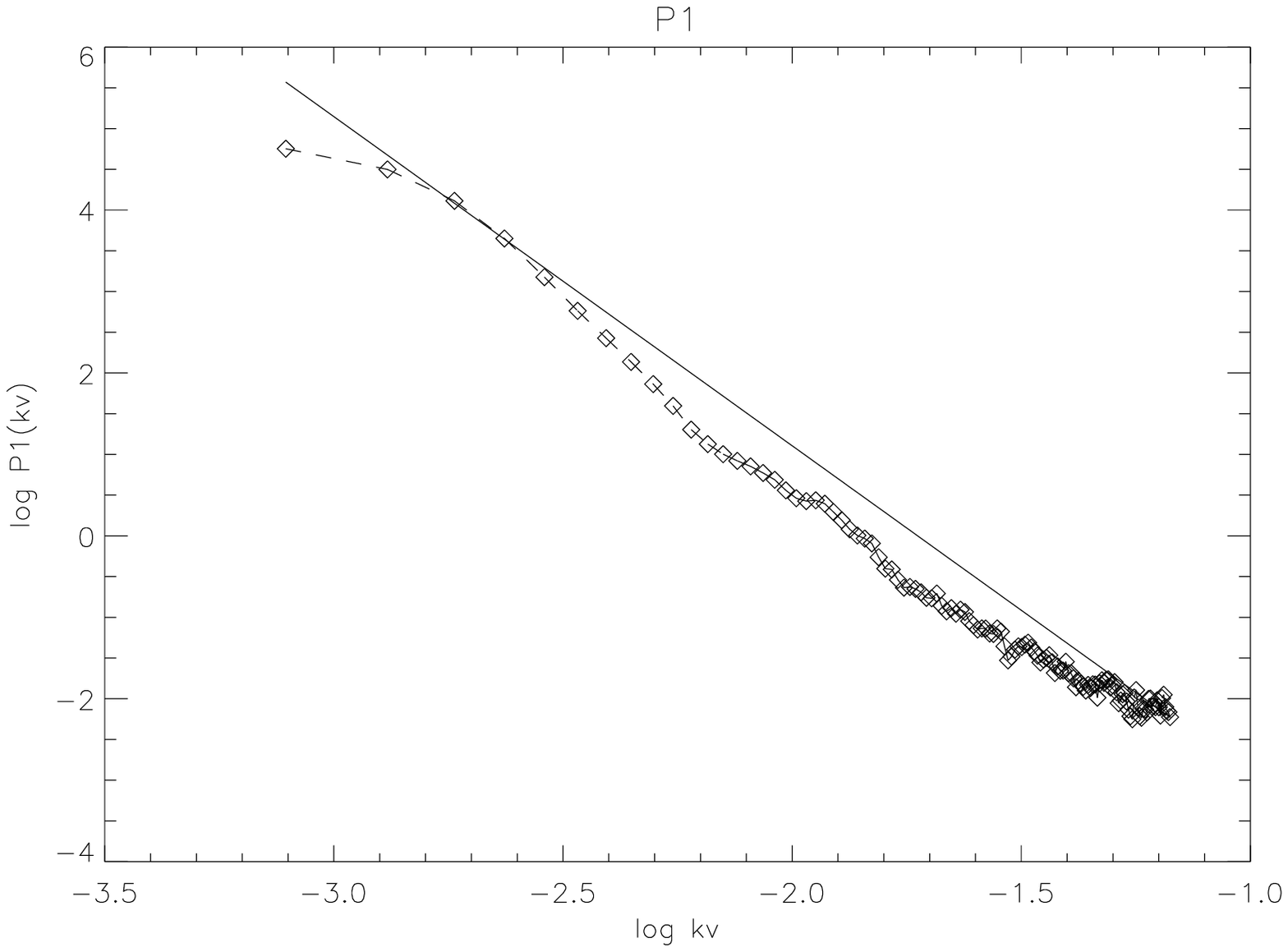}
\end{center}
\caption{2D VCS simulations for high-resolution (left) and low-resolution (right) modes. Solid lines show expected slopes. Velocity spectral index is $4$. \label{fig:VCS_4_00_2d}} 
\end{figure}

\begin{figure}
\begin{center}
\plotone{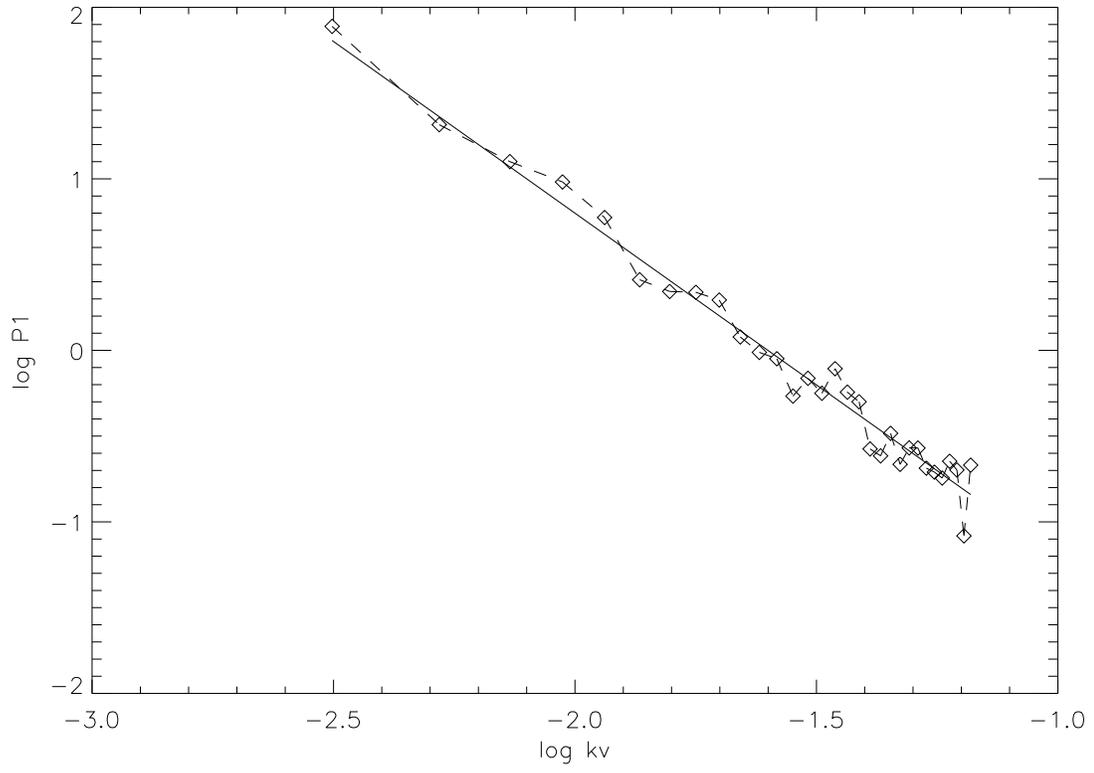}
\end{center}
\caption{VCS spectrum for 10 realizations in high-resolution mode. This illustrates the opportunity to recover velocity spectrum from absorption lines against a limited number of stars.
\label{fig:abs_10}} 
\end{figure}



\newpage

\begin{thebibliography}{}

\bibitem[Balesteros-Peredes et al. (2006)]{Bal06}
Ballesteros-Paredes, J.; Klessen, R. S.; Mac Low, M.-M.; Vazquez-Semadeni, E.
Protostars and Planets V, B. Reipurth, D. Jewitt, and K. Keil (eds.), University of Arizona Press, Tucson, 951 pp., 2007., p.63-80

\bibitem[Bereznyak, Lazarian \& Cho (2005)]{Ber05}
Beresnyak A., Lazarian, A., Cho, J., ApJ, (2005), 624, L93

\bibitem[Biskamp 2003]{Bis03}
Magnetohydrodynamic Turbulence, by Dieter Biskamp,Cambridge, UK: Cambridge University Press, 2003.

\bibitem[Chepurnov \& Lazarian, (2008)]{LC08}
Chepurnov, A., Lazarian, A., 2008, submitted to ApJ

\bibitem[Chepurnov et al. (2008)]{Che08}
Chepurnov, A., Lazarian, A., Stanimirovic, S., Peek, J. E. G., Heiles, C., (2008) in preparation 

\bibitem[Cho et al. (2003)]{Cho03}
Cho, J., Lazarian, A., \& Vishniac, E. T., 
ApJ, (2003), 564, pp. 291-301.

\bibitem[Cho$,$ Lazarian \& Vishniac, (2002)]{CLV02}
Cho, J., Lazarian, A., \& Vishniac, E. T., ApJ, (2002), 564, 291 

\bibitem[Cho \& Lazarian, (2002)]{CL02}
Cho, J., \& Lazarian, A., ApJ, (2002), 575, 63 

\bibitem[Cho \& Lazarian, (2006)]{ChL06}
Cho, J., Lazarian, A., ApJ, (2006, 638), 811.

\bibitem[Elmegreen, (2002)]{Elm02} Elmegreen, B., ApJ, (2002), 577, 206

\bibitem[Elmegreen \& Falgarone, (1996)]{EF96}
Elmegreen, B. \& Falgarone, E., ApJ, (1996), 471, 816

\bibitem[Elmegreen \& Scalo, (2004)]{ES04}
Elmegreen, B. \& Scalo, J., ARA\&A, (2004), 42, 211 

\bibitem[Ensslin \& Vogt, (2006)]{Ens06}
Ensslin, T. \& Vogt, C., A\&A, (2006), 453, 447

\bibitem[Esquivel et al. (2003)]{Esq03}
Esquivel, A., Lazarian, A., Pogosyan, D., Cho J., Mon. Not. R. Astron. Soc. (2003), 342, 325-336

\bibitem[Esquivel \& Lazarian, (2005)]{EL05}
Esquivel, A., Lazarian, A., ApJ, (2005), 320.

\bibitem[Falgarone et al. (2006)]{Fal06}
Falgarone, E., Pineau Des Gorets, B., Hily-Blant, P., \& Schilke, P., A\&A, (2006), 452, 511

\bibitem[Hartman \& McGregor, (1980)]{HMG80} 
Hartmann, L., \& MacGregor, K. B., ApJ, (1980), 242, 260

\bibitem[Inogamov \& Sunyaev, (2003)]{IS03}
Inogamov, N. A., Sunyaev, R. A., Astronomy Letters, (2003), 29, 791

\bibitem[Kim \& Ryu(2005)]{KR05} 
Kim, J., \& Ryu, D.\ 2005, \apjl, 630, L45 

\bibitem[Kowal et al.(2007)]{Kow07} 
Kowal, G., Lazarian, A., \& Beresnyak, A.\ 2007, \apj, 658, 423 

\bibitem[Lazarian, (2006)]{Laz06b}
Lazarian, A., AN, (2006), 327, 609

\bibitem[Lazarian, (2007)]{Laz06c}
Lazarian, A.,
2007 AAS/AAPT Joint Meeting, American Astronomical Society Meeting 209, 17.18

\bibitem{} Lazarian, A. 2008, Space Sci. Rev. in press

\bibitem[Lazarian \& Pogosyan, (2006)]{LP06}
Lazarian, A., Pogosyan, D., ApJ, (2006), 652, 1348

\bibitem[Lazarian \& Pogosyan, (2004)]{LP04}
Lazarian, A., Pogosyan, D., ApJ, (2004), 616, 943

\bibitem[Lazarian et al. (2001)]{Laz01}
Lazarian, A., Pogosyan, D., Vazquez-Semadeni, E., \& Pichardo, B., ApJ, 555, 130 (2001)

\bibitem[Lazarian \& Pogosyan, (2000)]{LP00}
Lazarian, A., Pogosyan, D., ApJ, (2000), 537, 720

\bibitem[Lazarian \& Vishniac, (1999)]{LV99}
Lazarian, A., \& Vishniac, E. ApJ, (1999), 517, 700

\bibitem[Lazarian$,$ Vishniac \& Cho, (2003)]{LVC03}
Lazarian, A., Vishniac, E., Cho, J., \& Ethan, T., ApJ, (2003), 595, 812

\bibitem[Lazarian \& Yan, (2002)]{LY02}
Lazarian, A. \& Yan, H. 2002, ApJ, 566, L105

\bibitem[Mac Low \& Klessen, (2004)]{MLK04}
Mac Low, M. \& Klessen, R., (2004), RvMP, 76, 125

\bibitem[McKee \& Tan, (2002)]{MKT02}
McKee, C., Tan, J., Nature, (2002), 416, 59

\bibitem[Miville-Deschênes et al. (2003)]{Miv03} 
Miville-Deschênes, M., Levrier, F., \& Falgarone, E. A\&A, (2003), 593, 831

\bibitem[Monin \& Yaglom (1975)]{MY}
Monin, A.S., Yaglom, A.M., 1975,
Statistical Fluid Mechanics: Mechanics of Turbulence (Cambridge: MIT Press)

\bibitem[Narayan \& Medvedev, (2001)]{NM01}
Narayan, R., \& Medvedev M.V., ApJ, (2001) 562, L129

\bibitem[Padoan et al. (2006)]{Pad06}
Padoan, P, Juvela, M., Kritsuk, A., \& Norman, M., 
ApJ, (2006), 653, pp. L125-L128.

\bibitem[Padoan et al. (2004)]{Pad04}
Padoan, P., Jimenez, R., Nordlund, A., Boldyrev, S.,
(2004), Physical Review Letters, 92, 19

\bibitem[Schlickeiser, (1999)]{Schlick99}
Schlickeiser, R., Quasilinear Theory of Cosmic Ray Transport in Weak Magnetohydrodynamic Plasma Turbulence,  
(1999), ed. M. Ostrowski and R. Schlickeiser, Cracow, 225

\bibitem[Stutzki, (2001)]{Stu01}
Stutzki, J. Astrophysics and Space Science Supplement, (2001), 277, 39

\bibitem[Sunyaev$,$ Norman \& Bryan, (2003)]{Sun03}
Sunyaev, R.A., Norman, M.L., \& Bryan, G.L. Astronomy Letters, (2003), 29, 783

\bibitem[Vazquez-Semadeni et al. (2000)]{Vaz00}
Vazquez-Semadeni, E., Ostriker, E. C., Passot, T., Gammie, C. \& Stone, J., 
in ``Protostars \& Planets IV'', eds. V. Mannings, A. Boss \& S. Russell (Tucson: Univ. of Arizona Press), 
(2000), 3 

\bibitem[Yan \& Lazarian, (2002)]{YL02}
Yan, H. \& Lazarian, A., Phys. Rev. Lett., (2002), 89, 28

\bibitem[Yan \& Lazarian, (2004)]{YL04}
Yan, H. \& Lazarian, A., ApJ, (2004), 614, 757

\end{thebibliography}
\end{document}